\documentclass[printer]{aa}
\usepackage{graphicx}
\usepackage{natbib}
\usepackage{txfonts}
\begin{document}
   \titlerunning{ACIS-I observations of NGC~2264}
   \title{ACIS-I observations of NGC~2264. Membership and X-ray
properties of PMS stars}

   \subtitle{}

   \author{E. Flaccomio
          \and
          G. Micela 
          \and
          S. Sciortino 
          }

   \offprints{E. Flaccomio}

   \institute{INAF - Osservatorio Astronomico di Palermo 
              Giuseppe S. Vaiana, 
	      Palazzo dei Normanni, 90134 Palermo, Italy \\
              \email{ettoref@astropa.inaf.it, giusi@astropa.inaf.it, 
	       sciorti@astropa.inaf.it }
             }

   \date{Received February --, 2006 accepted April 5,  2006}

   \abstract {}{Improving the member census of the NGC~2264 star
forming region and studying the origin of X-ray activity in young PMS
stars.}  {We analyze a deep, 100\ ksec long, Chandra ACIS observation
covering a 17'x17' field in NGC~2264. The preferential detection in
X-rays of low mass PMS stars gives strong indications of their
membership. We study X-ray activity as a function of stellar and
circumstellar characteristics by correlating the X-ray luminosities,
temperatures and absorptions with optical and near-infrared data from
the literature.} {We detect 420 X-ray point sources. Optical and NIR
counterparts are found in the literature for 85\% of the sources. We
argue that more than 90\% of these counterparts are NGC~2264 members,
thus significantly increasing the known low mass cluster population
by about 100 objects. Among the sources without counterpart about
50\% are likely associated with members, several of which we expect to
be previously unknown protostellar objects. With regard to activity we
confirm several previous findings: X-ray luminosity is related to
stellar mass, although with a large scatter; $L_X/L_{bol}$ is close to
but almost invariably below the saturation level, $10^{-3}$,
especially when considering the {{\em quiescent}} X-ray emission. A
comparison between CTTS and WTTS shows several differences: CTTS have,
at any given mass, activity levels that are both lower and more
scattered; emission from CTTS may also be more time variable and is on
average slightly harder than that of WTTS. However, we find evidence in
some CTTS of extremely cool $\sim0.1-0.2$keV, plasma which we speculate
is heated by accretion shocks.} {Activity in low mass PMS stars, while
generally similar to that of saturated MS stars, may be significantly
affected by mass accretion in several ways: accretion is likely
responsible for very soft X-ray emission directly produced in the
accretion shock; it may reduce the average energy output of solar-like
coronae, at the same time making them hotter and more dynamic. We
briefly speculate on a physical scenario that can explain these
observations.}


\maketitle 
%

\section{Introduction}

The collapse of molecular cores and the early evolution of pre-main
sequence (PMS) stars+disk systems involve a variety of complex
phenomena leading to the formation of main sequence (MS) stars and
planetary systems. Most of these phenomena, and their influence on the
outcome of the formation process, are not yet fully understood. 

X-ray observations of star forming regions have proved an invaluable
tool for star formation studies. On one hand, because of the much
higher luminosity of PMS stars in the X-ray band with respect to older
field stars, deep imaging observations are one of the few effective
means of selecting unbiased samples of members comprising both
classical T-Tauri stars (CTTS) and, most importantly, the otherwise
hard to distinguish weak line T-Tauri stars (WTTS). Selection of a
complete member sample is of paramount importance for any star
formation study, such as those focused on the initial mass function
\citep{salp55}, the star formation history \citep[e.g.][]{pal00}, the
evolution of circumstellar disks and planetary systems
\citep[e.g.][]{hai05}, and binarity \citep[e.g.][]{lada06}.  On the
other hand, the conspicuous X-ray activity of PMS stars is one of the
aspects of the PMS stellar evolution that are not yet well understood,
both with respect to its physical origin and to its consequences for
the stellar/planetary formation process. Indeed, the ionization and
heating caused by the penetrating X-ray emission might have a
significant impact on the evolution of star/disk systems
\citep{ige99,gla04} as well as on that of the star forming cloud as a
whole \citep{lor01}.

The high X-ray activity levels of PMS stars \citep[e.g. ][]{pre05} have
often been attributed to a ``scaled up'' solar-like corona formed by
active regions. This is the same picture proposed for MS stars, for
which the X-ray activity is related to the stellar rotation \citep[e.g.
][]{piz03}, evidence that a stellar dynamo is responsible for the
creation and heating of coronae. For most non-accreting PMS stars
(WTTS), the fractional X-ray luminosity, $L_{\rm X}/L_{\rm bol}$, is
indeed close to the saturation level, $10^{-3}$, seen on rapidly
rotating MS stars \citep{fla03b,pre05,piz03}. This might suggest a
common  physical mechanism for the emission of X-rays in WTTS and MS
stars or, at least, for its saturation. However, the analogy with the
Sun and MS stars may not be fully valid: first of all, the relation
between activity and rotation is not observed in the PMS
\citep{pre05,reb06}. Moreover, with respect to the Sun at the {\em
maximum} of its activity cycle, saturated WTTS have $L_{\rm X}/L_{\rm
bol}\sim1000$ times greater and plasma temperatures that are also
significantly higher. 

The X-ray emission of CTTS, PMS stars that are still undergoing mass
accretion, poses even more puzzles. With their circumstellar disks and
magnetically regulated matter inflows and outflows, CTTS are complex
systems. With respect to their X-ray activity, the bulk of the
observational evidence points toward phenomena similar to those 
occurring on WTTS. However, CTTS have significantly lower and
unsaturated values of $L_{\rm X}$ and $L_{\rm X}/L_{\rm bol}$
\citep{dami95,fla03a,fla03b,pre05}. In apparent contradiction with this
latter result, high-resolution X-ray spectra of two observed CTTS, TW
Hydrae \citep{kas02,ste04} and BP Tau \citep{sch05}, have indicated
that soft X-rays may be also produced in accretion shocks at the base
of magnetic funnels. Moreover, magnetic loops connecting the stellar
surface with the inner parts of a circumstellar disk may produce some
of the strongest and longer lasting flares observed on PMS stars
\citep{fav05}. The recent detection of X-ray rotational modulation
\citep{fla05}, however implies that emitting structures are generally
compact, so that these long loops cannot dominate the quiescent
X-ray emission.

NGC~2264 is a $\sim$3 Myr old Star Forming Region located at $\sim$760
pc \citep{sun97} in the Monoceros. Compared to the Orion Nebula
Cluster (ONC) and Taurus, NGC~2264 has intermediate stellar density and
total population, making it an interesting target for investigating the
dependence of star formation on the environment. It is on average older
than the ONC ($\sim 1$Myr), but star formation is still active inside
the molecular cloud in at least two sites where a number of protostars
and prestellar clumps have been detected \citep{you06,pere06}. It is
therefore an useful target for the study of the formation and time
evolution of young stars. Its study is eased by the presence of an
optically thick background cloud, effectively obscuring unrelated
background objects, and by the low and uniform extinction of the
foreground population  \citep{walker56,reb02}. Despite being the first
star forming region ever identified as such, the low mass population of
NGC~2264 is still not well characterized: proper motion studies
\citep{VSB} have been restricted to high mass objects; several studies
have identified the classical T-Tauri population using disk and
accretion indicators \citep{park00,reb02,lamm04}. Past X-ray
observations with {\em ROSAT} \citep{fla00} have been useful in
identifying the weak line T-Tauri (WTTS) population, but have not been
sensitive enough to detect low mass ($M\le0.3M_\odot$) and embedded
stars. We present here results from the analysis of a deep {\em
Chandra} observation of the region. Another similar observation of
a region just to the north of the one here considered has been analyzed
by \citet{ram04a}. The X-ray properties of NGC~2264 members derived in
the present paper and by \citet{ram04a}, augmented with similar data
for the Orion Flaking Fields \citep{ram04b} are studied by
\citet{reb06} in comparison with the results of the COUP survey
\citep{get05}. Results from the same {\em Chandra} observation analyzed
here on the peculiar binary system KH~15D have been presented by
\citet{herb05}. Finally, the properties of three embedded X-ray sources
near Allen's source, observed with {\em XMM-Newton}, have been recently
presented by \citet{sim05}.

The paper is organized as follows: we begin (\S \ref{sect:Xdata}) with
the presentation of the X-ray data, its reduction, source detection and
photon extraction. In \S \ref{sect:ancdata} we then introduce the
optical and near infrared data used to complement the X-ray
observation.  In \S \ref{sect:analysis} we present the temporal an
spectral analysis of X-ray sources and we derive X-ray luminosities.
Sections \ref{sect:memb} and \ref{sect:results} then discuss our
results with respect to cluster membership and the origin of X-ray
activity on PMS stars. We finally summarize and draw our conclusions in
\S \ref{sect:sumconc}.

\section{X-ray data and preparatory analysis}
\label{sect:Xdata}

\subsection{The ACIS-I observation}

\begin{figure}[t]
\centering
\includegraphics[width=8.8cm]{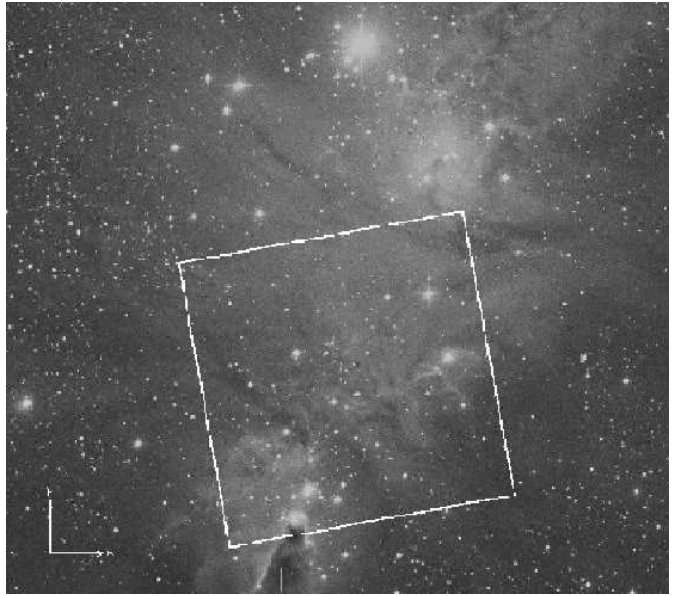}
\caption{Digitized Sky Survey image of NGC~2264. The field of view of
the {\em Chandra}-ACIS observation discussed in this paper 
is shown as a white square. The
famous Cone Nebula is visible toward the bottom of the image and 
the O7 star S Mon is close to the upper edge.}
\label{fig:acis_FOV}
\end{figure}

\begin{figure*}[t]
\centering
\includegraphics[width=17cm]{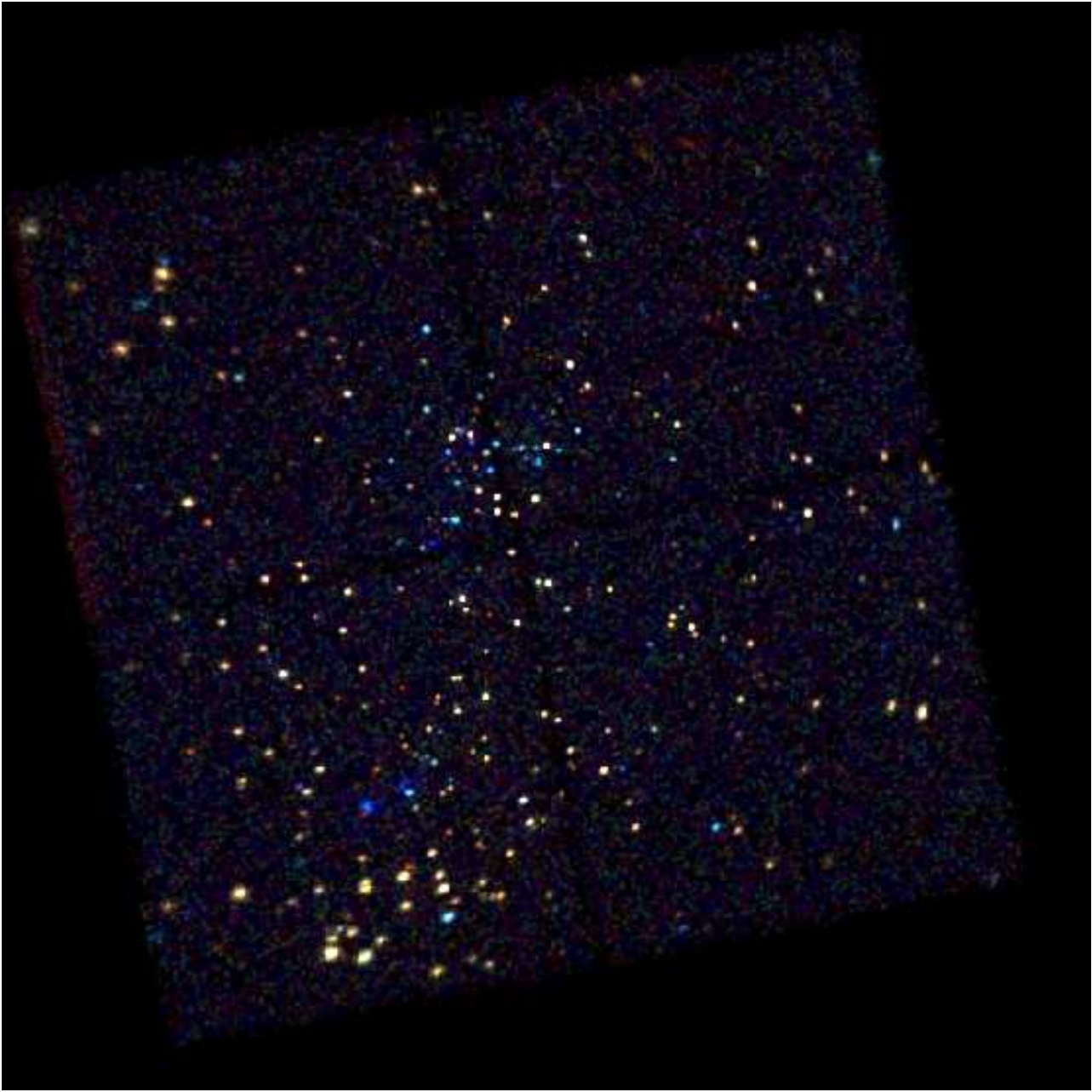}
\caption{NGC~2264 as seen in X-rays by ACIS. The true color (RGB)
image is
constructed from images in three energy bands: [200:1150]eV (red),
[1150:1900]
(green) and [1900:7000] (blue). Red therefore indicates soft and unabsorbed 
sources; blue hard and/or absorbed sources.}
\label{fig:acis_image}
\end{figure*}

We obtained a 97 ks long ACIS-I exposure of NGC~2264 on 28 Oct. 2002
(Obs. Id. 2540; GO proposal PI S. Sciortino). The 
17\arcmin$\times$17\arcmin field of view (FOV) of ACIS is shown in
figure \ref{fig:acis_FOV}, superimposed on the Digitized Sky Survey
image of the region. It was centered on R.A. 6$\rm ^h$ 40$\rm ^m$
58\fs7, Dec. 9\degr 34\arcmin 14\arcsec (roll angle: 79\degr). Figure
\ref{fig:acis_image} shows a color rendition of the spatial and
spectral information we obtained. ACIS was operated in FAINT mode with
CCD 0, 1, 2, 3, 6 and 7 turned on. Data obtained with CCD 6 and 7, part
of the ACIS-S array, will not be discussed in the following because of
the much degraded point spread function (PSF) and effective area
resulting from their large distance from the optical axis.

\subsection{Data preparation}

Data reduction, starting from the level 1 event file,  was performed in
a standard fashion, using the CIAO 2.3 package and following the
threads provided by the {\em Chandra X-ray Center}\footnote{see
cxc.harvard.edu}. Several IDL custom programs were also employed.
First, we corrected the degradation in the spectral response due to the
Charge Transfer Inefficiency (CTI), occurred in particular during the
first months of the {\em Chandra} mission using the {\sc
acis\_process\_events} CIAO task.  We then produced a level 2 event
file by retaining only events with grade=0,2,3,4,6 and status=0.
Finally we corrected the data for the time dependence of the energy
gain using the {\sc corr\_tgain} utility.  

X-ray stellar sources have, on average, a different spectrum with
respect to the ACIS background. The total signal to noise ratio (SNR)
of sources can therefore be maximized by filtering out events with
energy outside a suitable spectral band: we first performed a
preliminary source detection as discussed in \S \ref{sect:srcdet} on
the whole event list. We then defined for each source a radius,
$R_{97}$, such that 97\% of the PSF counts fall within this radius (\S
\ref{sect:phext}). We extracted source photons for all sources from
circles with $R=0.5\times R_{97}$ and background photons from a single
background region that excludes photons from all sources within their
respective $R_{97}$ (a sort of ``Swiss cheese'' image with the sources
carved out). We then computed the total SNR of
sources\footnote{$SNR_{tot}=\sum{(S^i- \alpha_i B)}/\sqrt{\sum{(S^i+
\alpha_i B)}}$ where $\alpha_i=A^i_s/A_b$; $S_i$ and $A^i_s$ are the
number of counts in the $i$-th source regions and their area
respectively. $B$ and $A_b$ are the background counts and relative
extraction area, taken for the present purpose as uniform within the
FOV.} for a fine grid of minimum and maximum energy cuts. The highest
source SNR was obtained for $E_{min}=200eV$ and $E_{max}=7000eV$. With
these cuts the number of photons in the source extraction area
(including background photons) is reduced to 96\% of the total, while
the background is reduced to 28\% of the total. We checked that
consistent results are obtained by maximizing the SNR of faint sources
only ($<$20 net counts), which may have a different average spectrum
and are the ones we are most interested in for the purpose of
detection.

After filtering in energy, the time integrated background is 0.07
counts per arcsec$^2$, consistent with nominal values. The background
was constant in time except for a small flare with a peak reaching
about twice the quiescent rate. The flare starts 13 ks after the
beginning of the observation and lasts $\sim 1$ ks. The effect of the
background flare on source light curves is however negligible, even for
faint sources, i.e. the most affected by the background. This is
confirmed by the negative results of Kolmogorov-Smirnov variability tests
(see \S \ref{sect:vari}) performed on the background extraction regions
relative to each source. In the study of source lightcurves (\S
\ref{sect:vari}) we will therefore assume a constant background.

\subsection{Source detection}
\label{sect:srcdet}

We detected sources using the PWDetect code
\citep{dam97}.\footnote{Available at: \\ {
http://www.astropa.unipa.it/progetti\_ricerca/PWDetect}} The
significance threshold was set to 4.6$\sigma$. According to extensive
simulations of source-free fields with the background level of our
observation, this threshold corresponds to an expectancy of 10 spurious
sources in the whole FOV. PWDetect reports 423 sources. Upon careful
inspection we removed three entries relative to sources that were
detected twice, leaving a total of 420 distinct sources. Twenty eight
of these are below the 5.0$\sigma$ significance threshold,
corresponding to the more conservative criteria of one expected
spurious source in the FOV. Background subtracted source counts, in the
0.2-7 keV band, are derived by PWDetect directly from the wavelet
transform of the data. Effective exposure times at the source
positions, averaged over the PSF, are also computed by PWdetect from an
exposure map created with standard CIAO tools assuming an input energy
of 2.0 keV\footnote{The choice of energy is not crucial: the ratio
between the effective area at the source position and on the optical
axis, which is the quantity we use to define the effective exposure
time, has only a small dependence on energy.}.  

Detected sources are listed in Table \ref{tab:xsrc}. In the first eight
columns we report source number, sky positions with uncertainty,
distance from the {\em Chandra} optical axis, source net counts (in the
0.2-7keV band), effective exposure time, and the statistical
significance of the detection. 

\begin{table*}
\caption{Catalog of X-ray ACIS detections. \label{tab:xsrc}}
\begin{center}
\begin{tabular}{ccccccccccccc}
\hline\hline
N & RA & Dec. & $\delta_{r,d}$ & Offax & Net Cts. & Exp. T. & Signif & $P_{KS}$ & $n_H$ & $n_H$(Ref) & Mod & $F_X$[u] \\
 & [h m s] & [d m s]. & ["] & ['] &  & [s] &  &  & [$10^{22}$cm$^{-2}$] &  &  & [$ergs/s/cm^2$] \\
\hline\hline
  1&   6:40:25.8&   9:29:24.0&      3.14&       9.5&      25.1&     80579&       6.1&     0.088&      1.59&    JHK&      --
&      -14.03\\
  2&   6:40:27.7&   9:32:00.1&      1.52&       8.0&      81.2&     82888&      16.2&     0.262&      0.00&     Av&      1T
&      -14.35\\
  3&   6:40:28.1&   9:35:33.7&      1.37&       7.7&      44.7&     81810&       9.0&     0.032&      0.00&     Av&      --
&      -14.52\\
  4&   6:40:28.6&   9:35:47.3&      1.22&       7.6&     186.6&     64151&      29.6&     0.000&      0.10&     Av&      1T
&      -13.69\\
  5&   6:40:28.8&   9:30:59.6&      0.59&       8.1&    1222.2&     82889&      70.2&     0.000&      0.26&     Av&      2T
&      -12.74\\
  6&   6:40:28.8&   9:33:05.3&      0.80&       7.5&     111.0&     83404&      17.7&     0.579&      0.00&     Av&      1T
&      -14.12\\
  7&   6:40:30.6&   9:38:37.7&      1.84&       8.2&      11.1&     82255&       4.7&     0.724&        --&     --&      --
&          --\\
  8&   6:40:30.9&   9:34:40.4&      0.92&       6.9&     179.1&     83958&      25.7&     0.001&      0.62&      X&      1T
&      -13.48\\
  9&   6:40:31.2&   9:31:07.0&      0.71&       7.5&     506.4&     84042&      45.7&     0.093&      0.09&     Av&      2T
&      -13.31\\
 10&   6:40:31.7&   9:33:28.9&      1.64&       6.7&      18.9&     84683&       6.0&     0.045&      0.00&    JHK&      --
&      -14.91\\
\hline
\end{tabular}
\end{center}
First 10 rows. The entire table, containing 420 entries, is available in the electronic edition of A\&A.
\end{table*}

\subsection{Photon Extraction}
\label{sect:phext}

Source and background photon extraction for spectral and timing
analysis was performed using CIAO and custom IDL software. We first
determined for each source the expected PSF using the CIAO {\sc mkpsf}
tool, assuming a monochromatic source spectrum (E=1.5 keV). We  thus
determined the expected encircled energy fraction as a function of
distance from the source. Photons extraction circles were defined so as
to contain 90\% of the PSF, save for 28 sources for which the encircled
PSF fractions were reduced to values ranging from 74\% to 89\% to avoid
overlap with neighboring sources. The local background for each source
was determined from annuli whose inner radii excludes 97\% of the PSF
and with outer radii twice as large. In order to exclude contamination
of the background regions from the emission of neighboring sources we
excluded from these annuli photons falling within the 97\% encircled
counts radii of all sources. The area of background extraction regions
were computed through a mask in which we {\em drilled} regions outside
the detector boundaries and circles containing 97\% of the PSF photons
from all sources.

For sources with at least 50 photons, source and background spectra
suited to the XSPEC spectral fitting package were then created using
standard CIAO tools.  Corresponding response matrices and effective
areas (RMF and ARF files respectively) were also produced with CIAO.
For spectral analysis, spectra were energy binned so that each bin
contains a fixed number of photons, depending on net source counts
$N_{src}$ : 15 photons per bin for $N_{src}>200$, 10 photons for
$100<N_{src}<200$ and 5 photons for $N_{src}<100$. Because spectral
analysis (\S \ref{sect:spec}) was restricted to energies $>0.5$keV, the
first energy bin was forced to begin at that energy.

\section{Ancillary data}
\label{sect:ancdata}

\subsection{Cross-identifications - optical/NIR catalogs}
\label{sect:crossid}

We have cross-identified our X-ray source list with catalogs from the
following optical/NIR surveys covering the whole area of our ACIS
field: 2Mass (NIR photometry), \citet[optical photometry +
spectroscopy]{walker56}, \citet[optical/NIR photometry + low resolution
optical spectroscopy]{reb02}\footnote{The \citet{reb02}, catalog was
updated following private communication from L. Rebull}, \citet[optical
photometry + variability]{lamm04}, \citet[optical photometry +
spectroscopy]{dahm05}, \citet[X-ray sources]{fla00}.  The seven
cross-identified catalogs are listed in the first column of table
\ref{tab:catalogs}. In the 2nd column we indicate the number of objects
within the ACIS FOV and in the third the number of objects identified
with ACIS sources and of those for which the identification is unique.
Adopted positional tolerances for cross-identifications are reported in
the 4th column, either as a single figure for the whole catalog or as a
range when defined for each individual object (see the table
footnotes). They are based on the uncertainties quoted for each
catalog.

\begin{figure}[t]
\centering
\includegraphics[width=8.9cm]{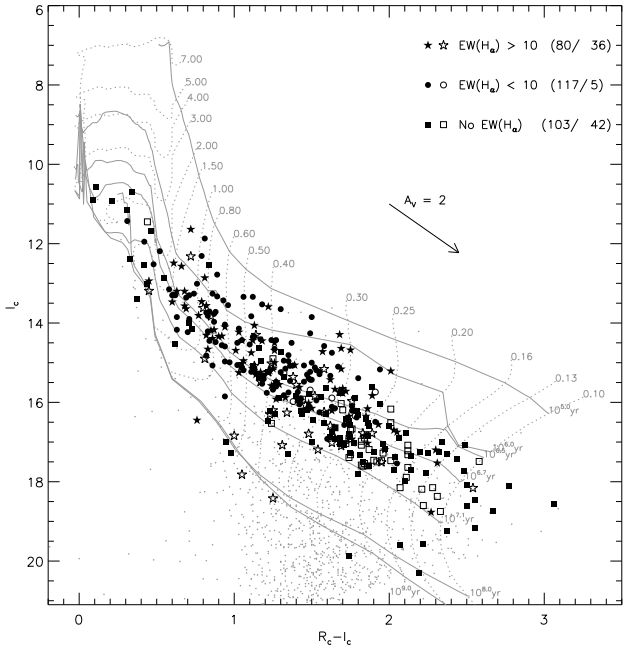}
\caption{Optical color magnitude diagram of all the objects in the ACIS
field of view. Larger symbols indicate X-ray sources or likely NGC~2264
members as defined in \S \ref{sect:starchar}. Filled symbols refer
to ACIS detected
objects. As indicated in the legend, when possible we distinguish 
between CTTS and WTTS as defined by the EW of their $H_\alpha$ line.
The number of detected and undetected objects in each class is also indicated
in the legend. The grid shows the SDF evolutionary tracks and isochrones
transformed to colors and magnitudes using the conversion table in
\citet{kh95}.}
\label{fig:cmdO}
\end{figure}

\begin{figure}[t]
\centering
\includegraphics[width=8.9cm]{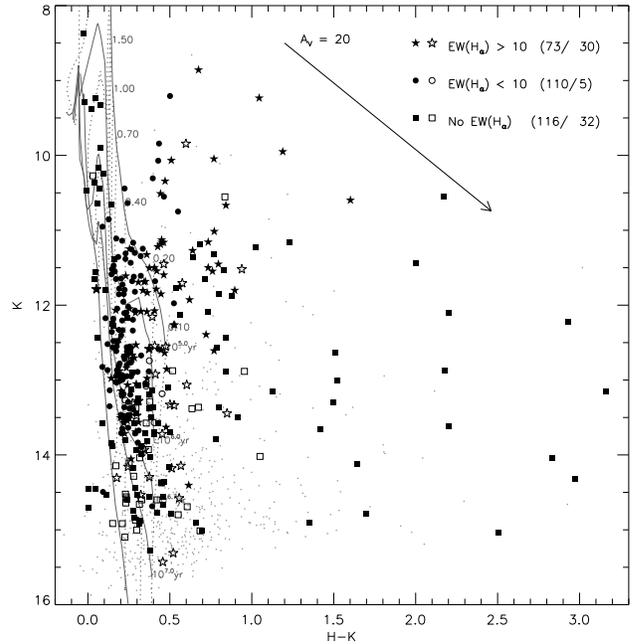}
\caption{Near-IR color magnitude diagram for the objects in the ACIS
FOV. Symbols and tracks as in figure \ref{fig:cmdO}.}
\label{fig:cmdIR}
\end{figure}

Among the catalogs here considered, the deepest photometric surveys are
2Mass in the NIR (JHKs\footnote{Note that in the following we
neglect the small differences among NIR photometric systems, and in
particular that between the 2MASS Ks and  the standard K bands.}) and
\citet{lamm04} in the optical (V$R_c$$I_c$). A comparison with the
isochrones of \citet[][hereafter, SDF]{sie00} in the optical and NIR
color-magnitude diagrams (Fig. \ref{fig:cmdO} and \ref{fig:cmdIR}),
indicates that at the distance of NGC~2264, 2Mass reaches down to about
0.1$M_\odot$ for 10Myr old stars, i.e. the oldest expected in the
region, while \citet{lamm04} reaches slightly deeper for unabsorbed
stars, but is obviously less sensitive to highly absorbed ones. Within
the ACIS FOV, spectral types are given for 7 stars by \citet{walker56},
for 87 stars by \citet{reb02}, for 150 stars by \citet{lamm04} and for
157 stars by \citet{dahm05}.

\begin{table*}
\caption{Catalogs used for cross-identification\label{tab:catalogs}}
\begin{center}
\begin{tabular}{l r r c c c c} 
\hline\hline
Catalog           & $N_{obj}$&  $N_{det}/N_{unique}$   & Id.Rad.	& $\Delta$R.A.& $\Delta$Dec.& $\rm Off_{50\%} $\\
                  &          &  	    & [\arcsec]   &   [\arcsec] & [\arcsec]   & [\arcsec]    \\
\hline
  ACIS  	  &	 420 &  ---	    & 1.0-12.6$^a$& -0.09	& -0.09        &  0.24 \\
  2Mass 	  &	1098 &  346/344	    &  0.5-1.4$^b$&  0.00	&  0.00        &    -- \\
  \citet{walker56}&	  67 &    52/52	    &	  1.0	  & -0.45	&  0.08        &  0.47 \\
  \citet{reb02}   &	 511 &  236/235	    &	   1.5    &  0.73	& -0.03        &  0.73 \\
  \citet{lamm04}  &	1598 &  305/299	    &	   1.0    & -0.13	&  0.45        &  0.47 \\
  \citet{dahm05}  &	 229 &  183/183	    &	   1.0    & -0.16	&  0.18        &  0.36 \\
  \citet{fla00}   &	  84 &    80/80	    & 8.1-42.8$^c$& -0.85	&  1.49        &  3.70 \\
\hline 
\end{tabular}
\end{center}
$^a$ $\max[2\sigma_{rd},1\arcsec]$. Mean/median: 1.74\arcsec/1.25\arcsec \\
$^b$ $\max[2(\sigma_{R.A.}^2+\sigma_{Dec.}^2)^{1/2},0.5\arcsec]$.
Mean/median: 0.54\arcsec/0.50\arcsec \\
$^c$ 85\% point source encircled energy radii \citep{fla00}.
Mean/median: 12.0\arcsec/9.2\arcsec.

\end{table*}

We created a master list of cross-identified objects following a
step-by-step procedure. In the first step we matched ACIS sources with
2Mass objects: first we registered the ACIS coordinates to the 2Mass
ones by iteratively cross-identifying the two catalogs and shifting the
ACIS coordinates by the mean offset of uniquely identified source
pairs. Identification radii were chosen as the quadrature sum of the
above defined position tolerances (table \ref{tab:catalogs}). We then
created a joint catalog of objects containing all matched and unmatched
ACIS and 2Mass objects, assigning to them coordinated from 2MASS when
available. In the following step we repeated the above process using
the ACIS+2Mass catalog as reference and matching it with the
\citet{reb02} one. We then repeated the process with the \citet{fla00}
X-ray source list, then with the \citet{lamm04}, \citet{dahm05} 
and \citet{walker56} catalogs. The coordinate shifts of all considered
catalogs with respect to the 2Mass system are given in table
\ref{tab:catalogs} (columns 5 and 6), along with the median offsets
between object positions and the reference 2Mass catalog. After each
step, identified source pairs and unidentified objects were checked
individually and a small fractions of the identifications ($<1\%$) were
modified. In the first step for example, five identifications between
four ACIS and five 2Mass sources were added: in two cases (sources \#64
and \#404) 2Mass sources were only slightly more distant with respect
to the identification radii.  In another case the X-ray source, \#102,
was situated at the edge of the detector and its positions was surely
more uncertain than the formal error indicated. In the last case the
X-ray source, \#237, was detected between two close-by 2Mass objects
and an identification was forced with both.

Table \ref{tab:master} lists, for the 1888 distinct objects in the ACIS
FOV, consolidated coordinates and cross-identifications numbers for
each of the seven catalogs. 425 rows refer to objects related to one of
the 420  ACIS sources: 351 are identified with a single optical/NIR
counterpart, two are identified with 5 and 2 counterparts respectively,
and 67 do not have any optical or near-infrared identification. The
other 1463 rows in table  \ref{tab:master} refer to non ACIS-detected
objects. For these latter we computed upper limits to the ACIS count
rate using PWdetect and the same event file used for source detection.
Measured count rates, repeated from table \ref{tab:xsrc}, and upper
limits are reported in column 12 of table \ref{tab:master}.

Focusing on the identification of ACIS sources with optical/NIR
catalogs, given the relatively large number objects in the field of
view, we can wonder how many of the identifications are due to chance
alignment and not to a true physical association. We can constrain the
number of chance identifications by assuming that positions in the two
lists are fully uncorrelated. Because this is surely not the case for
our full X-ray source catalog, our estimate can only be considered a
loose upper limit. Furthermore, limiting our X-ray sample to the 28
sources with significance below 5.0$\sigma$, 9 of which are expected to
be spurious and whose positions will indeed be random, we can place an
upper limit on the fraction of spurious sources associated with an
optical object. This value is of interest when studying the X-ray
properties of optically/NIR selected samples. In order to estimate the
number of spurious identifications assuming uncorrelated positions, we
proceed as follows: for each X-ray source we consider optical objects
within a circular neighborhood of area  $A_{nei}$, within which source
density is assumed uniform; we then estimate the fraction of $A_{nei}$
covered by identification circles. The sum of these fractions is our
upper limit to the number of chance identifications. We repeated the
calculation for radii of the  neighborhood circle from 1.0\arcmin\ to
4.0\arcmin. The upper limit to the number of chance identifications for
the full sample of ACIS sources ranges from 13 to 15. For the 28 source
with  significance $<5.0\sigma$ we instead estimate no more than 2.5
chance identifications. If we assume that out of the 28 sources, only
the nine spurious detections have positions that are indeed
uncorrelated with optical objects, we can scale this result and
conclude that $\sim 1$ spurious X-ray source will be by chance
identified with an optical object. Conversely, we can most likely
locate spurious sources among the 15 sources with significance below
5.0$\sigma$ and that are not identified with optical/NIR catalogs.

\begin{table*}
\caption{Master catalogs  of objects in the ACIS FOV.\label{tab:master}}
\begin{center}
\begin{tabular}{lccccccccccc}
\hline\hline
N & RA & Dec. & Id. rad. & ACIS & 2Mass & Reb.+ & Lamm+ & Flacc+ & W56 & Dahm+ & Ct. Rate. \\
 & [h m s] & [d m s] & ["] &  &  &  &  &  &  &  & [$10^{-4}s^{-1}]$ \\
\hline\hline
   1 &   6:40:22.4&   9:28:03.2&      0.54&     --    &    06402243+0928032    &     --    &   3228    &     --    &     --    &     --    &$<$    8.7
\\
   2 &   6:40:24.2&   9:30:38.1&      0.50&     --    &    06402419+0930381    &   2229    &     --    &     --    &     --    &     --    &$<$    3.3
\\
   3 &   6:40:24.3&   9:27:44.4&      0.50&     --    &    06402425+0927443    &     --    &     --    &     --    &     --    &     --    &$<$   23.5
\\
   4 &   6:40:24.8&   9:29:39.6&      0.66&     --    &    06402475+0929395    &     --    &     --    &     --    &     --    &     --    &$<$    3.4
\\
   5 &   6:40:24.8&   9:28:42.6&      0.50&     --    &    06402482+0928425    &     --    &     --    &     --    &     --    &     --    &$<$    3.3
\\
   6 &   6:40:24.8&   9:28:59.8&      0.64&     --    &    06402482+0928597    &     --    &     --    &     --    &     --    &     --    &$<$    3.5
\\
   7 &   6:40:25.0&   9:30:00.5&      1.00&     --    &                  --    &     --    &   3295    &     --    &     --    &     --    &$<$    3.0
\\
   8*&   6:40:25.0&   9:32:08.4&      0.50&     --    &    06402504+0932084    &   2258    &   3299    &     --    &     --    &     --    &$<$    4.1
\\
   9 &   6:40:25.1&   9:29:36.9&      0.50&     --    &    06402505+0929369    &   2261    &   3300    &     --    &     --    &     --    &$<$    3.4
\\
  10 &   6:40:25.4&   9:30:26.4&      1.00&     --    &                  --    &     --    &   3326    &     --    &     --    &     --    &$<$    2.8
\\
\\
\hline
\end{tabular}
\end{center}
*: Likely NGC~2264 member;\\  \dag: Entry also present in another row.\\
First 10 rows. The entire table, containing 1888 entries, is available in the electronic edition of A\&A.
\end{table*}

\begin{figure}[t]
\centering
\includegraphics[width=8.9cm]{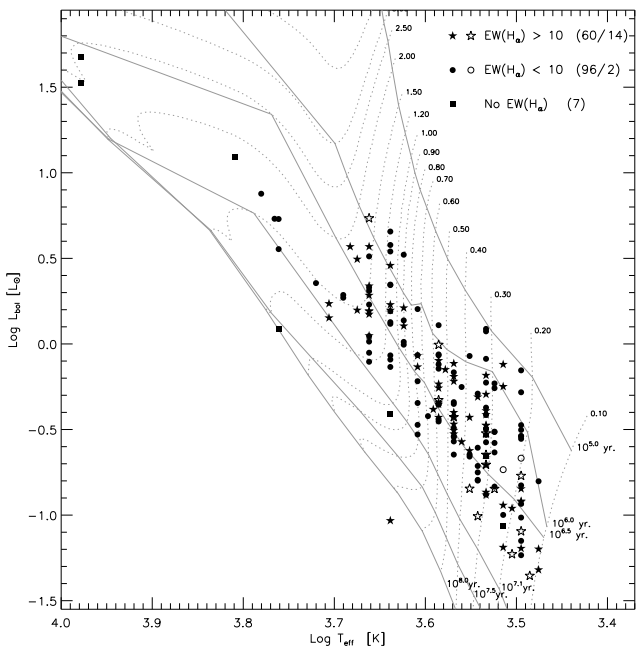}
\caption{Theoretical HR diagram for the subsample of likely members that
could be placed in this diagram trough optical photometry and
spectral types. Symbols as in figure \ref{fig:cmdO}. Tracks and
isochrones are by SDF.}
\label{fig:HR}
\end{figure}

\subsection{Characterization of stars in the FOV}
\label{sect:starchar}

We have collected optical and NIR data from the literature for all the
stars in the master catalog assembled in the previous section. In table
\ref{tab:xopt} we report photometry and spectral types for our X-ray
sources with unique optical identification. A total of 300 X-ray
detected stars have been assigned both $R_c$ and $I_c$ magnitudes and
are plotted in figure \ref{fig:cmdO} as filled symbols; 299 have H and
K magnitudes and are plotted in figure \ref{fig:cmdIR} (264 of these
also appear in Fig. \ref{fig:cmdO}). In both color magnitude diagrams
(CMDs) we also show for reference SDF tracks and isochrones,
transformed to colors and magnitudes using the conversion
table given by \citet{kh95} and shifted along the reddening vectors by
the median extinction of known members ($A_V=0.44$) and
vertically by the distance modulus corresponding to the adopted
distance. For the extinction law we adopted \citet{rie85}. For
stars with spectral types we derived effective temperatures, $T_{eff}$,
bolometric corrections, $BC_I$, and intrinsic colors, $(R-I)_0$, using
the relations compiled by \citet{kh95} and, for the temperature of M
stars, the intermediate gravity scale of \citet{luh99}. Using the
available $R_c$ and $I_c$ photometry we then derived extinction values
($A_V= 4.46\times E(R-I)$, where $E(R-I)=(R-I)_0-(R-I)$) and bolometric
luminosities ($L_{bol}= -0.4\times[I_c-BC_I-A_V/1.63-DM(760pc)$]).
Finally we estimated masses and ages from the theoretical HR diagram,
figure \ref{fig:HR}, through interpolation of the SDF evolutionary
tracks. In summary, out of the 351 X-ray sources with an unique
optical/NIR identification, we estimated $T_{eff}$ for 165 X-ray
sources, $A_V$ and $L_{bol}$ for 163, masses and ages for 161.

Other than the sample of X-ray detected stars, that, as discussed in \S
\ref{sect:memb} are likely cluster members, we will also consider
another sample of 83 X-ray undetected likely cluster members. These
latter, plotted with empty symbols in figure \ref{fig:cmdO},
\ref{fig:cmdIR} and  \ref{fig:HR}, are chosen according to their
position in the I vs. R-I diagram, the strength of the $H_\alpha$ line
(measured either spectroscopically or photometrically) and optical
variability: first, we define a {\em cluster locus} in the I vs. R-I
diagram using the SDF tracks and the observed concentration of X-ray
sources. The cluster locus is defined as the area above the  $10^{7.1}$
Myr isochrone {\em or} to the left of the 0.8$M_\odot$ evolutionary
track.\footnote{This latter condition is required by the convergence of
the isochrones at higher masses and by the observed spread of stars in
the ``cluster locus'', at least in part due to uncertainties.} We then
consider as likely members: stars in the cluster locus and with strong
$H_\alpha$ emission according to the narrow-band $H_\alpha$ photometry
of \citet{lamm04}, using the same criterion discussed by these authors;
stars in the cluster locus with moderate (chromospheric) $H_\alpha$
emission and with variable optical lightcurves (both periodic and
irregular), again as discussed by \citet{lamm04}; stars for which
spectroscopic observations of the  $H_\alpha$ line are available and
for which the measured EW is larger than 10 (the canonical CTTS
threshold). Among this sample of 83 X-ray undetected likely members we
are able to determine masses and ages for 16 stars.

\begin{table*}
\caption{Optical/NIR properties of X-ray sources. \label{tab:xopt}}
\begin{center}
\begin{tabular}{cccccccccccccc}
\hline\hline
N & R & R-I & J & H & K & Sp.T. & Av & Av(JHK) & $\log T_{eff}$ & $\log L_{bol}$ & Mass & LogAge & $P_{rot}$ \\
 &  &  &  &  &  &  &  &  & [K] & [$L_\odot$] & [$M_\odot$] & [yr.] & [d] \\
\hline\hline
  1&     17.97&      2.10&     15.92&     14.17&     13.37&        --&        --&      9.96&        --&        --&        --&        --&        --\\
  2&     15.68&      1.25&     14.27&     13.54&     13.29&      M2.5&      0.00&      0.62&      3.54&     -0.80&      0.34&      6.63&      1.23\\
  3&     15.88&      1.33&     14.44&     13.70&     13.50&        M3&      0.00&      0.00&      3.53&     -0.87&      0.30&      6.63&     12.09\\
  4&     14.14&      1.03&     12.79&     12.07&     11.85&        M0&      0.62&      0.41&      3.59&     -0.06&      0.57&      6.11&      4.55\\
  5&     13.26&      0.86&     11.81&     11.30&     10.75&        K4&      1.61&      0.00&      3.66&      0.51&      1.57&      6.19&      7.49\\
  6&     15.15&      1.33&     13.66&     12.97&     12.78&        M3&      0.00&      0.00&      3.53&     -0.57&      0.32&      6.33&      0.76\\
  7&     19.86&      1.74&        --&        --&        --&        --&        --&        --&        --&        --&        --&        --&        --\\
  8&     14.39&      0.86&     12.69&     11.51&     10.67&        K1&      1.92&      2.39&      3.71&      0.15&      1.30&      7.10&     12.09\\
  9&     13.82&      0.83&     12.57&     11.90&     11.62&        K6&      0.58&      0.00&      3.62&      0.01&      0.92&      6.38&      1.98\\
 10&     17.06&      1.70&     15.42&     14.85&     14.60&        --&        --&      0.00&        --&        --&        --&        --&        --\\
\hline
\end{tabular}
\end{center}
First 10 rows. The entire table, containing 420 entries, is available in the electronic edition of A\&A.
\end{table*}

In order to derive intrinsic X-ray luminosities of detected stars or
upper limits for undetected ones, knowledge of the extinction is
fundamental (\S \ref{sect:Xlum}). For stars with no optical spectral
type $A_V$ cannot be determined as indicated above.  We therefore
estimated extinctions from the NIR J-H vs. H-K diagram by de-reddening
stars that could be placed in this diagram onto the expected intrinsic
locus. This latter was taken from \citet{kh95} for the stellar
contribution and supplemented with the CTTS locus of \citet{mey97}. The
degeneracy in the reddening solutions was solved by always taking the
first intercept between the intrinsic locus and the reddening vector.
Stars whose position was within 1.5$\sigma$ of the loci were assigned
zero extinction. In summary, we obtained a total of 278 and 62
extinction values for X-ray detected stars and undetected likely
members, respectively. These number exceed by 115 and 46 the number of
extinctions derived from optical data for the same samples. Although
these NIR extinction values are surely more uncertain than the ones
derived from spectral types, they are especially valuable for highly
extinct objects.

\subsection{Cross-identifications - MIR/mm catalogs}
\label{sect:crid_mir}

In addition to the wide field surveys described in \S
\ref{sect:crossid} we have also correlated our X-ray sources with three
mid-infrared (MIR) and millimeter (mm) catalogs recently published for
the IRS~2 and IRS~1 regions: \citet[][hereafter Y06]{you06},
\citet[][T06]{teix06} and \citet[][P06]{pere06}. These three surveys
target very young and/or embedded objects and are therefore ideal for
checking the nature of X-ray sources with no optical/NIR counterpart.
They are however limited in area coverage: T06 lists Class I/0 sources
detected with {\em SPITZER} in a  $\sim3\times 3$\ arcmin$^2$ region close
to IRS~2; Y06 lists all {\em SPITZER} detected objects in a dense but
even smaller $\sim 2\times 2$\ arcmin$^2$ region slightly south-east of
IRS~2; P06 list the pre/protostellar cores detected at 1.2mm in both
the IRS~1 and IRS~2 regions. 

Within the area of Y06, the only MIR work available so far that lists
all the objects detected in the surveyed region, we find MIR
counterparts for 2 out of 4 X-ray sources lacking optical/NIR
counterparts. In total 4 of the 67 unidentified X-ray sources were
assigned new counterparts:  \#145 was associated with source 12 of T06,
a Class I/0 source;  \#228 was associated with the D-MM~15 mm core,
indicated by P06 as probably starless;  \#244 was associated with
source 43 in Y06 and is, judging from its spectral energy distribution
(SED), an absorbed class II/III PMS star that is relatively bright in
$K$-band ($K=12.15$, Y06)\footnote{Note that the source is missed by
2MASS because, at the 2MASS spatial resolution, it is blended with a
nearby source}; finally \#274 was associated with the D-MM~10 mm core,
indicated as starless by P06, with source  1656 in Y06 and source 15 in
T06 (characterized by a steeply rising MIR SED).

Among the X-ray sources with optical/NIR counterparts, 17 in the region
covered by Y06 were identified with {\em SPITZER} sources of which one,
\#281, is a likely Class I source (Y06's source 62). Three more X-ray
sources were identified with Class I/0 sources in the list of T06:
\#150 and \#242 also having NIR counterparts and \#194 (see also
\S \ref{sect:XMM}) with both NIR and optical ($I=19.89$) counterpart.
As for the P06 mm-cores, identifications are made uncertain by the
limited spatial resolution of the mm data (cf. P06). The core D-MM~14,
classified as protostellar by P06, is likely associated with ACIS
source \#89 (offset: 1.2''), an optically faint ($V=20.91$) and NIR
bright ($K=10.6$) star with large $H_\alpha$ emission:
$EW(H_\alpha)=458$. Quite similarly, C-MM~1, also indicated as a
protostellar core by P06, is offset by 1.3'' from the ACIS source
\#361, an optically visible source with strong $H_\alpha$ ($V=18.55$,
$K=11.85$, $EW(H_\alpha)=231$)\footnote{Although Y06 report that C-MM~1
has no NIR counterpart, we associate it with 2MASS source
06411792+0929011 (offset=1.5'').} and it is therefore likely associated
with it. Finally, C-MM~5, classified by P06 as a protostellar core,
might be associated with ACIS \# 305 (offset: 1.9''). This ACIS source
is in turn most likely associated with IRS~1 \citep[offset: 0.8'',
][]{schr03}. Note however that P06 suggests that C-MM~5 may not be
associated with IRS~1.


A more detailed analysis of the X-ray properties of young protostars in
NGC~2264 will be the subject of a future paper.

\section{Analysis}
\label{sect:analysis}

\subsection{Temporal analysis}
\label{sect:vari}

Source variability was characterized through the Kolmogorov-Smirnov
test. Column (9) of table \ref{tab:xsrc} reports the resulting
probability that the distribution of photon arrival times is not
compatible with a constant count-rate. Given the sample size (420
sources) a value below 0.1\% (obtained for 72 sources) indicates the
light-curve is almost certainly variable, while a $<$1\% value (87
sources) indicates a very likely variability, although up to $\sim 4
(=420\times 0.01)$ of the {\em variable} sources might actually be
constant. If we are not interested in the individual sources but on the
overall fraction of variable sources, we can place a lower limit on
this quantity by computing the minimum number of variable sources as a
function of probability threshold, $P_{th}$: $N_{min}= N(P_{KS} <
P_{th})-420\times P_{th}$. The maximum $N_{min}$ is obtained for
$P_{th}=15$\% for which $N(P_{KS} < P_{th})=189$ and $N_{min}=126$. We
conclude that within our observation  {\em at least} 30\% (126/420) of
the lightcurves are statistically inconsistent with constancy. This
fraction is surely a lower limit to the true number of variable
sources, as our ability to tell a variable source from a constant one
is influenced by photon statistics. This is clearly indicated by the
dependence of the fraction of variable sources on source counts. If for
example, we restrict the above analysis to sources with more than
100(500) counts, a total of 145(33) sources, we find that at least
54\%(73\%) of these are variable. 

\begin{figure}[!t!]
\centering
\includegraphics[width=8.5cm]{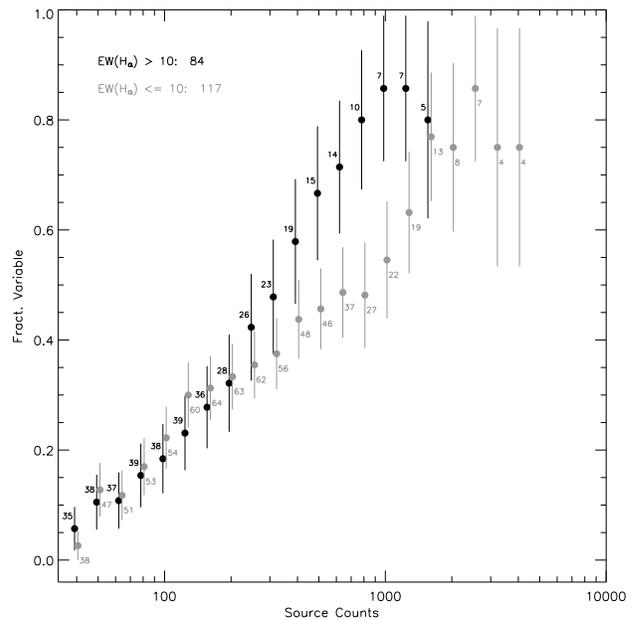}
\caption{Fraction of variable stars as a function of count statistics for
stars with $H_\alpha$ EW greater and smaller than 10 (i.e. CTTS and
WTTS). Note that counts bin are centered at the plot symbol and are 0.8 dex
wide, so that successive bins are not independent. Error bars are based
on binomial statistics. Symbols referring to the two subsamples are
slightly shifted in the horizontal
direction  with respect to each other so to avoid confusion. The total
number of objects in the subsample is given in the legend, while the
numbers of objects from which each variability fraction is computed are shown
beside the plotting symbols.}
\label{fig:var_fract_ha}
\end{figure}

\begin{figure}[!t!]
\centering
\includegraphics[width=8.5cm]{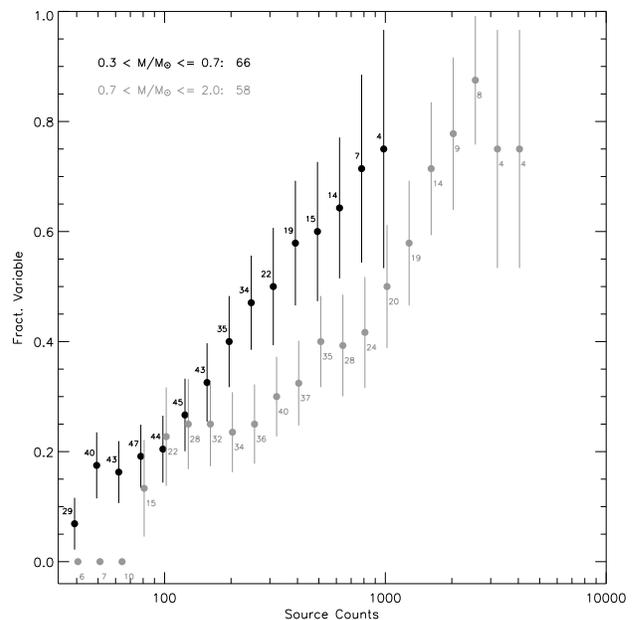}
\caption{Fraction of variable stars as a function of count statistics for
stars in two different mass ranges: [0.3-0.7]$M_\odot$ and
[0.7-2.0]$M_\odot$. Symbols and legend as in figure \ref{fig:var_fract_ha}.}
\label{fig:var_fract_mass}
\end{figure}

We next investigated whether different kind of stars, classified from
optical data, showed different variability fractions. From the previous
discussion, it is clear that for a meaningful comparison, source
statistic must be taken into account. Figure \ref{fig:var_fract_ha}
shows the variability fraction, $f(P_{KS}<1\%)$, as a function of
source counts for CTTS and WTTS, as discriminated by their $H_\alpha$
equivalent width.  Variability fractions are computed for sources with
counts in intervals spanning 0.8 dex (because in the figure points are
spaced by 0.1 dex, only one point every eighth is independent). Error
bars on the variability fractions are estimated assuming binomial
statistics: $\delta f= [f\times(1-f)/N_{src}]^{1/2}$, where $N_{src}$
is the number of sources in each count bin. In addition to the expected
increase of the variability fraction with source statistics, we note
that CTTS appear to be more variable with respect to WTTS, at least
when considering stars with more than $\sim$200 counts. Figure
\ref{fig:var_fract_mass} shows a similar comparison for two mass
segregated sub-samples: $0.3<  M/M_\odot < 0.7$ and $0.7 <  M/M_\odot <
2.0$. Lower mass stars appear to be more variable, and again this
difference is noticeable only for stars that are bright enough. We
quantified the differences between variability fractions, by testing,
for each count bin, the null hypothesis that the two samples are drawn
from the same parent population. We chose as statistics the difference
of the two observed variability fractions, $\Delta f$. We then
numerically computed the probability that a $\Delta f$ equal or larger
than the observed one is obtained by randomly drawing pairs of numbers
from binomial distributions appropriate to the two sample sizes but
characterized by the same probability of observing a variable source,
i.e the variability fraction. Because this latter number is not well
constrained we conservatively took as our confidence level the minimum
that is obtained varying this fraction between 0 and 1. The results of
these test indicate that the differences in the variability fractions
are statistically not highly significant: in the bin centered at $\sim
400$ counts, the low mass stars are more variable than higher mass
stars with a 94\% confidence. In the bin centered at $\sim 800$ counts,
CTTS are more variable than WTTS with a 92\% confidence. We thus
consider these results as tentative. However we note that the
significance of the difference between CTTS and WTTS is strengthened by
the fact that a similar result, and with a similar confidence, was also
obtained by \citet{fla00} using totally independent ROSAT data.

\subsection{Spectral analysis}
\label{sect:spec}

We have analyzed the X-ray spectra of the 199 sources with more than 50
detected photons. Spectral fits were performed with XSPEC 11.3
and several shell and TCL scripts to automate the process. For each
source we fit the data in the [0.5-7.0]keV energy interval\footnote{
Events with energies below 0.5keV were not used because of
uncertainties in the calibration resulting from the time dependent
degradation of the ACIS quantum efficiency at low energies.} with
several model spectra: one and two isothermal components ({\sc apec}),
subject to photoelectric absorption from interstellar and circumstellar
material ({\sc wabs}). Plasma abundances for one temperature models
were fixed at 0.3 times the solar abundances \citep{wilms00}, while for
two temperature models they were both fixed at that value and treated
as a free parameter. The absorbing column densities, $N_H$, were both
left as a free parameter and fixed at values corresponding to the
optically/NIR determined extinctions, when available: $N_H=1.6\
10^{21}A_V$ \citep{vuo03}. A total of three or six models were thus fit
for each source depending on the availability of optical extinction
values. For each model, spectral fits were performed starting from
several initial conditions for the fit parameters as indicated in table
\ref{tab:models}. For example for isothermal models with free $N_H$,
two values of $N_H$ and four values of $kT$ were adopted as initial
conditions, for a total of eight distinct fits. For each model the
adopted fit parameter set was chosen from the model fit that minimize
the $\chi^2$. This procedure was adopted in order to reduce the risk
that the $\chi^2$ minimization algorithm used by XSPEC finds a relative
minimum.

\begin{table}
\caption[]{Initial conditions for XSPEC models.}
\label{tab:models}
\begin{tabular}{l c c c}
  \hline
  Model name &  $N_H$ [cm$^{-2}$]                & $kT_1/kT_2$                 & Abund.\\
  \hline
   1T   & 0.0, $10^{22}$ &    0.5/--, 1.0/--, 2.0/--, 10.0/--       & {\em 0.3} \\
   2T   & 0.0, $10^{22}$ &    0.4/1.0, 1.0/3.0, 0.5/2.0	& {\em 0.3} \\ 
   2Tab & 0.0, $10^{22}$ &    0.4/1.0, 1.0/3.0, 0.5/2.0	& 0.3 \\ 
   1T$_{Av}$  & {\em 1.6$\times$10}$^{21}A_V$ 	 &    0.5/--, 1.0/--, 2.0/--, 10.0/--        & {\em 0.3} \\
   2T$_{Av}$   & {\em 1.6$\times$10}$^{21}A_V$   &    0.4/1.0, 1.0/3.0, 0.5/2.0 & {\em 0.3} \\
   2Tab$_{Av}$ & {\em 1.6$\times$10}$^{21}A_V$ 	 &    0.4/1.0, 1.0/3.0, 0.5/2.0	& 0.3 \\ 
  \hline
\end{tabular} \\
Multiple values indicate that all combinations of initial values were
adopted.
Values in italic indicate that fit parameters were fixed to those
values.
\end{table}

\begin{figure}[t]
\centering
\includegraphics[width=8.5cm]{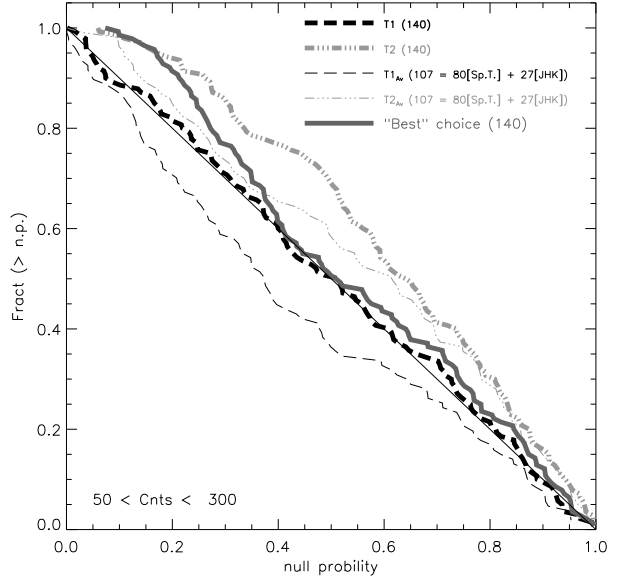}
\caption{Cumulative distribution of null probabilities resulting from X-ray
 spectral fits. The curves refer here to X-ray sources with more that
50 counts (the minimum for which we performed spectral fits) and less 
than 300 counts. The different lines refer to different physical models
(one and two temperature, with free of fixed $N_H$)
as indicated in the legend, along with the number of sources that enter
in each distribution. For the models with fixed $N_H$ we also indicate
the number of $N_H$ derived from optical spectral types + photometry and
of those derived from 2MASS J, H and K photometry. The gray thick line
refers to the ``best'' choice of models described in \S \ref{sect:spec}.}
\label{fig:spec_np_dist_a}
\end{figure}

\begin{figure}[t]
\centering
\includegraphics[width=8.5cm]{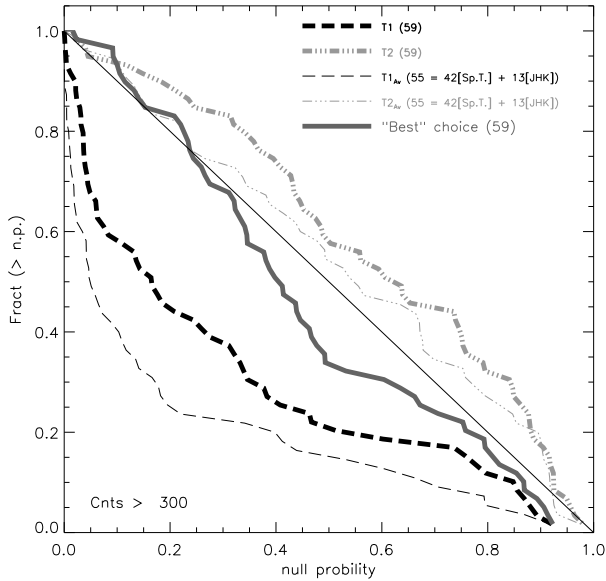}
\caption{Same as figure \ref{fig:spec_np_dist_a} for sources with more
than 300 counts.}
\label{fig:spec_np_dist_b}
\end{figure}

Next, we considered which of the available model fits for each source
(three or six, reference model names are given in table
\ref{tab:models}) was the most representative of the true source
spectra, and thus the one to be adopted for the following
considerations. The goal was twofold: to characterize the emitting
plasma in order to investigate its origin, and to determine accurate
intrinsic band-integrated X-ray luminosities. Crucial for this latter
step is the determination of extinction ($N_H$, see \S
\ref{sect:Xlum}). Generally speaking, models must have enough
components to yield statistically acceptable fits according to the
$\chi^2$ or, equivalently, the null probabilities (n.p.) that the
observed spectra are described by the models. However, models with too
many free parameters with respect to the spectra statistics, while
formally yielding good fits, will not be constrained by the data and
will yield limited physical information. A particularly severe problem
with CCD quality (ACIS) low statistic spectra is the degeneracy between
absorption and temperature: an equally good fit can be often obtained
with a cool plasma model with a large emission measure but suffering
high absorption, or with a warmer temperature and a lower extinction.
It is therefore desirable to check the $N_H$ obtained from the fits
with independent information from optical/NIR data. Figures
\ref{fig:spec_np_dist_a} and \ref{fig:spec_np_dist_b} show, for faint
and bright sources ($50 < cnts < 300$, $cnts > 300$) respectively, the
cumulative distribution of the n.p. for spectral fits performed with
four different models. The models are 1T, 2T, and, for sources with
independent $N_H$ estimates, 1T$_{Av}$ and 2T$_{Av}$. In this kind of
plot the distribution for a perfectly adequate spectral model should
follow the diagonal, that for an oversimplified model should fall below
the diagonal, and that for an over-specified model should lie above.
For faint sources 1T model appears perfectly adequate. Fixing the $N_H$
to the optically determined value worsen the agreement somewhat but
still results in fits that are, when considered individually, for the
most part acceptable. A two temperature model with free $N_H$ appears
to be too complicated (and therefore unconstrained), while a 2T model
with fixed $N_H$ is more acceptable, but still on average too
sophisticated for the quality of these low-statistic spectra. For
brighter sources we notice instead that 1 temperature models are on
average not favored, while 2T models are still not always needed.

For most sources more than one spectral model is statistically
acceptable. We choose a best guess model as the simplest that still
gives a statistically acceptable fit. The compelling reason to choose
this approach is the mentioned degeneracy between temperature and
absorption. Note for example that from figure \ref{fig:spec_np_dist_a},
one could be tempted to always choose, for faint sources, 1T models
with free $N_H$, as these models appear to be statistically perfectly
adequate to represent the spectra. However, when examined individually,
many of these 1T spectral fits have rather degenerate fit solutions,
i.e. large and correlated uncertainties on the $N_H$ and $kT$ values,
also implying large uncertainties on unabsorbed fluxes. This 
degeneracy can be broken by using the additional information on
absorption coming from optical/NIR data, when available and compatible
with the X-ray spectra. 

After some experimenting, we choose our best guess model according to
the following empirical scheme: if n.p.(1T$_{A_v}) > 20\%$ we chose the
1T$_{Av}$ model. Otherwise, if n.p.(1T)$ > 20\%$ we chose the 1T model
(with free $N_H$). If both of the previous tests for isothermal models
failed, we tried with two temperature models, this time lowering our
n.p. threshold to 5\%. First we tried the 2T$_{Av}$ model
(n.p.(2T$_{A_v}) > 5\%$) and then the 2T model (n.p.(2T)$ > 5\%$). As a
last resort, in case none of the previous models could be adopted, we
choose the 2Tab model (free $N_H$, free abundances). Although the above
scheme was designed so to favor simple models, the different n.p.
thresholds resulted in four cases (sources \#257, \#275, \#280 and
\#300) in adopting 2Tab models that were either statistically
worse or comparable to simpler 1T, 1T$_{Av}$ or 2Tab$_{Av}$
models. We therefore adopted these latter. After careful examination of
individual fits, the spectral models adopted for four more sources
(\#97, \#104, \#127 and \#241) were modified. In these cases the
automatic choice would result in unphysical, unusual, or unconstrained
temperatures and absorptions, whereas our adopted models are acceptable
both statistically and physically.\footnote{For source \#241 we changed
the 2T model (n.p.=32\%, $kT_1=0.057^{7.8}_{0.022}$keV, 
$kT_2=0.57^{1.30}_{0.30}$keV, $N_H=13^{21}_{1.8}\ 10^{21}\ cm^{-2}$) 
to a  1T model (n.p.=17\%, $kT=0.60^{0.84}_{0.38}$keV,
$N_H=3.3^{6.5}_{0.0}\ 10^{21}\ cm^{-2}$). For source \#104 we changed
the 1T model (n.p.=40\%, $kT=0.43^{0.73}_{0.29}$keV,
$N_H=3.0^{5.6}_{0.78}\ 10^{21}\ cm^{-2}$) to a  $2T_{Av}$ model
(n.p.=32\%, $N_H(A_V)=0.0 cm^{-2}$ and reasonable temperatures). For
sources \#97 and \#127 we changed from $1T_{Av}$ models (p=68\% \&
p=0.21\%, $kT=54$ keV \&  $kT=19$ keV, both with unconstrained
uncertainties) to 1T models (p=97\% \& p=56\%,  $kT=4.7$ keV \&
$kT=3.5$ keV). }

The solid gray lines in figure \ref{fig:spec_np_dist_a} and
\ref{fig:spec_np_dist_b} refer to the final choice of best guess
models. Table \ref{tab:xspec} lists the end result of our spectral
analysis: we report the adopted model, the source of the adopted $N_H$,
the null probability, the plasma temperature(s) and normalization(s),
the observed and absorption corrected fluxes in the [0.5-7]keV band. 

\begin{table*}
\caption{Spectral properties of ACIS  sources with more than 50 counts. \label{tab:xspec}}
\begin{center}
\begin{tabular}{ccccccccccc}
\hline\hline
N & Mod & $n_H$(Ref) & $P_{null}$ & $n_H$ & $kT_1$ & $n_1$ & $kT_2$ & $n_2$ & $F_X$[a] & $F_X$[u] \\
 &  &  &  & [$10^{22}$cm$^{-2}$] & [keV] &  & [keV] &  & [$ergs/s/cm^2$] & [$ergs/s/cm^2$] \\
\hline\hline
  2&      1T&     Av&      0.40&$      0.00_{       }^{       }$&$      0.86_{   0.72}^{   1.07}$&$     -5.23_{  -5.35}^{  -5.14}$&$          --_{       }^{       }$&$        --_{       }^{       }$&      -14.35&      -14.35\\
  4&      1T&     Av&      0.62&$      0.10_{       }^{       }$&$      1.29_{   1.12}^{   1.63}$&$     -4.58_{  -4.65}^{  -4.53}$&$          --_{       }^{       }$&$        --_{       }^{       }$&      -13.78&      -13.69\\
  5&      2T&     Av&      0.14&$      0.26_{       }^{       }$&$      0.26_{   0.22}^{   0.30}$&$     -3.63_{  -3.77}^{  -3.51}$&$        1.23_{   1.15}^{   1.34}$&$     -3.88_{  -3.92}^{  -3.84}$&      -13.07&      -12.74\\
  6&      1T&     Av&      0.85&$      0.00_{       }^{       }$&$      1.13_{   0.92}^{   1.39}$&$     -5.00_{  -5.09}^{  -4.93}$&$          --_{       }^{       }$&$        --_{       }^{       }$&      -14.12&      -14.12\\
  8&      1T&      X&      0.41&$      0.62_{   0.36}^{   0.99}$&$      9.64_{   3.36}^{  64.00}$&$     -4.67_{  -4.76}^{  -4.49}$&$          --_{       }^{       }$&$        --_{       }^{       }$&      -13.62&      -13.48\\
  9&      2T&     Av&      0.92&$      0.09_{       }^{       }$&$      0.31_{   0.25}^{   0.40}$&$     -4.32_{  -4.51}^{  -4.19}$&$        1.26_{   1.12}^{   1.53}$&$     -4.40_{  -4.48}^{  -4.34}$&      -13.43&      -13.31\\
 11&      1T&    JHK&      0.17&$      0.10_{       }^{       }$&$      0.98_{   0.86}^{   1.13}$&$     -4.63_{  -4.70}^{  -4.57}$&$          --_{       }^{       }$&$        --_{       }^{       }$&      -13.86&      -13.75\\
\hline
\end{tabular}
\end{center}
First 7 rows. The entire table, containing 199 entries, is available in the electronic edition of A\&A.
\end{table*}

In summary, out of the 199 sources with more than 50 counts, we adopt
isothermal models in 147 cases and two component models in the
remaining 52 cases. For 138 sources we adopt a spectral fit in which
the $N_H$ was fixed, in 101 cases to the value estimated from the
optically determined $A_V$ and in 37 cases from NIR photometry. In the
remaining 61 cases $N_H$ was treated as a free fit parameter. Out of
these 61 cases, independent estimates of extinction were available from
optical and NIR data in 21 and 3 cases respectively, but our algorithm
(or, for sources \#97 and \#127, our choice) preferred the spectral fit
with free $N_H$. We examine these cases more in detail to assess the
ambiguities in the model fits and the consequences on the derived X-ray
fluxes.

\begin{figure}[t]
\centering
\includegraphics[width=8.9cm]{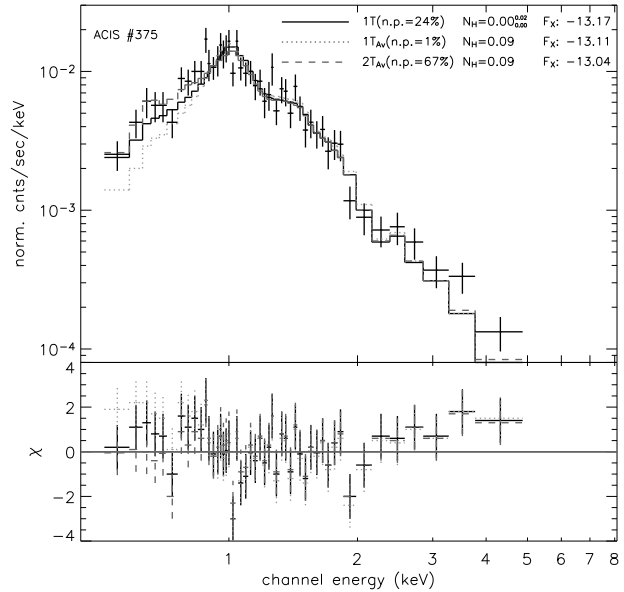}
\caption{Upper panel: spectrum of source \#375, with overlaid three different
spectral models, 1T with both free and fixed $N_H$ (solid and dotted
lines respectively) and 2T with
fixed $N_H$ (dashed line). Null probabilities, $N_H$ values (in units of
$10^{22}$ cm$^{-1}$) and unabsorbed 
fluxes for
each model are given in the legend. Note that the 1T and $2T_{Av}$
models are both statistically acceptable. Lower panel: residuals for the
three models in terms of sigmas (error bars all of unit size)}
\label{fig:spec_nh_ne_av_a}
\end{figure}

Figure \ref{fig:spec_nh_ne_av_a} illustrates one of such cases: source
\#375. Here the extinction determined from the 1T (kT=1.3keV) spectral
fit (90\% confidence interval: $N_H<2\ 10^{20}$cm$^{-2}$) is lower than
that estimated from the $A_V$ ($9\ 10^{20}$cm$^{-2}$). Fixing the $N_H$
to that value yields an unsatisfactory isothermal fit. We note however
that adding to the spectral model a second cool isothermal component
($kT=0.25$kev), a good fit can be obtained even fixing the $N_H$. This
is a quite typical example of the degeneracy in the fitting of sources
with low/moderate statistics. We can estimate the uncertainty in the
X-ray unabsorbed fluxes derived from the spectral fits due to this
degeneracy as the difference between the fluxes derived from the two
acceptable fits: for source \#375 for example this is $\sim 0.13$dex.
More in general, acceptable fits with $N_H$ fixed to the optical/NIR
values could be obtains in 22 of the 24 cases in which our procedure
preferred the spectral fit with free $N_H$. In five
cases\footnote{Sources \#190, \#340, \#352, \#375, \#391}, among which
that described above, in order to conciliate the observed spectra with
the optical extinction, an additional cool thermal component
(kT=0.22-0.34) would be required to compensate for the higher $N_H$. If
for these five sources the true $N_H$ were the optically derived ones,
by choosing the 1T fit with free (lower) $N_H$ we are underestimating
the unabsorbed fluxes by 0.14-0.37dex (mean 0.2dex). For the other 17
sources, fixing the $N_H$ would not require an additional component:
reasonable fits (n.p.$>$ 5\%) were obtained even with the same models
(1T in 16 cases, 2Tab for source \#187). In 10 of these 17 cases the
adopted fits with free $N_H$ result in smaller absorptions and slightly
hotter (0.2-0.3keV) temperatures (the cool temperature for \#187),
while the opposite happens in the other 7 cases. Had we fixed the $N_H$
to the optical values in these 17 cases we would have obtained
unabsorbed fluxes on average $\sim$0.03dex larger and always within
0.20dex (in either direction) of the adopted ones. 

From this discussion of the 22 cases with most ambiguous extinctions
($\sim 14\%$ of the sources for which we have optical/NIR $N_H$), we
conclude that: a) typical uncertainties in the unabsorbed flux are less
than 0.2dex, b) individual X-ray fluxes corrected using an $N_H$ from
spectral fits can be as wrong as $\sim$0.4 dex, if the cool temperature
is altogether missed by the spectral fit with a consequent
underestimation of the $N_H$.

\begin{figure}[t]
\centering
\includegraphics[width=8.9cm]{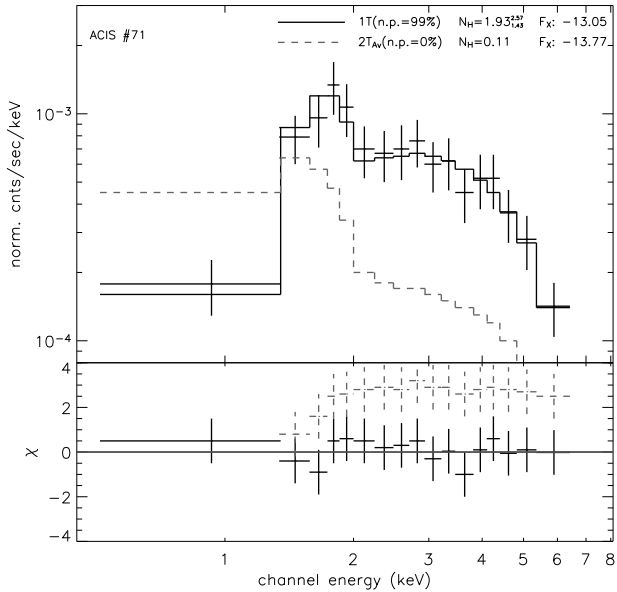}
\caption{Some as figure \ref{fig:spec_nh_ne_av_a} for source \#71. In
this case we plot the 1T model with free $N_H$ and the 2T model with
$N_H$ fixed to the optical value. Note how even adding a second
component, it is impossible to reconcile the observed spectrum with the
$N_H$ implied by the optical absorption and a standard absorption law.}
\label{fig:spec_nh_ne_av_b}
\end{figure}

Two more sources have incompatible X-ray and optical extinctions
according to our choice of {\em best} model. Source \#234 did not enter
in the previous discussion because it has the least acceptable fits of
the whole sample, at most n.p. =1.6\% for a 2T model with free $N_H$
(note, however, that with an adequate spectral model, we would expect
three sources out of 199 to have lower n.p.). Adopting the optical
$N_H$ ($1.8\times 10^{21}$) would result in an unabsorbed flux only
0.07dex larger with respect to that obtained leaving $N_H$ as a free
fit parameter ($N_H=0.59^{0.78}_{0.16}\times 10^{21}$). Source \#71 is
a more interesting case. Figure  \ref{fig:spec_nh_ne_av_b} shows its
spectrum, with spectral fits obtained with free and fixed $N_H$. The
X-ray derived extinction ($N_H=19^{26}_{14}\times 10^{21}$) is about 17
times larger than that estimated from the optical reddening
($N_H=1.1\times 10^{21}$)\footnote{For this source we have adopted the
spectral type (K2) from \citet{dahm05} and the photometry
($R_c-I_c=0.62$) from \citet{lamm04}. Independent photometry and
spectral types are available from \citet{reb02}, \citet{lamm04} and
\citet{dahm05}. $R_c-I_c$ ranges between 0.55 and 0.62 and spectral
types between G9 and K2. The possible range in $N_H =1.6\ 10^{21}A_V$
is thus $0.57-1.64\ 10^{21}$, i.e. between $\sim 9$ and $\sim 46$ times
the value derived from the X-ray spectrum, considering its 90\%
confidence interval.}. The discrepancy is highly significant and
independent of the considered spectral model. We note that the detected
emission from this  source is dominated by a powerful flare, with peak
count rate $\sim$100 times that before and after the flare. A possible
scenario to explain the high absorption might involve a solar-like
coronal mass ejection associated with the flare, providing the
additional absorbing material. Such an hypothesis has been formulated by
\citet{fav99} to explain a more modest increase in $N_H$ during a
powerful flare observed on Algol.

Further discussion of physical interpretation of spectral fit results
is deferred to \S \ref{sect:results}.

\subsection{X-ray luminosities}
\label{sect:Xlum}

\begin{figure}[t]
\centering
\includegraphics[width=8.5cm]{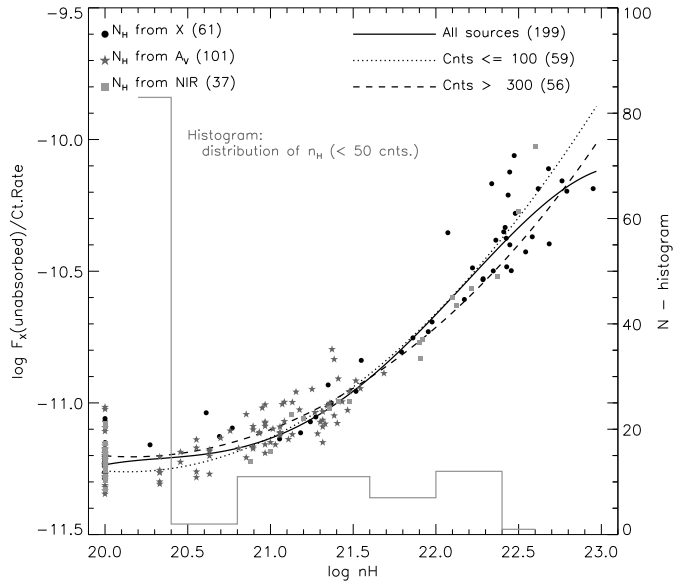}
\caption{Count-rate to unabsorbed flux conversion factor as a function of
$N_H$ as derived from spectral fits. Various symbols refer to
spectral fits in which $N_H$ was treated as a free parameter, fixed
to the value derived from the optical $A_V$ or to that determined
from the NIR photometry. The three curves indicate polynomial fits to the data
obtained for the whole sample of 199 sources and for subsamples of
sources with less than 100 counts (59 sources) or more than 300 counts
(56 sources). The gray histogram (vertical scale on the right hand side)
shows the distribution of absorption values, derived from the $A_V$ or
the NIR CCD, of X-ray sources with less than 50 counts, for which
count-rate to flux conversion factor was determined using the
polynomial fit to faint sources ($<$100 counts, dotted curve).}
\label{fig:cf}
\end{figure}

Extinction corrected X-ray luminosities in the [0.5-7.0]keV band were
estimated for all sources for which an indication of extinction was
available. For the 199 sources with more than 50 counts, and for which
spectral fits were performed, we computed $L_X$ from the $F_X[u]$
(unabsorbed fluxes) column in table \ref{tab:xspec}, adopting a
distance of 760pc, adequate for NGC~2264 members. For the fainter
sources, for which spectral fits were not performed, we derived a
count-rate to unabsorbed flux conversion factors using the results of
the spectral fits for the brighter stars. Figure \ref{fig:cf} shows,
for these latter sample, the run of the flux/rate ratio vs. $N_H$.
Different symbols indicate different origins of the absorption values
(see previous section). We then performed polynomial fits to the data
points for the whole sample, for the 56 sources with more than 300
counts and for the 59 sources with less 100 counts. Results are shown
in fig. \ref{fig:cf}. Because the sources for which we want to
determine fluxes have $<$50 counts, we adopt the latter fit as our
relation between $N_H$ and the conversion factor:

\begin{equation}
\log \frac{F_X(u)}{Ct.Rate}= 58.62 -6.942\times \log N_H+ 0.1724\times (\log
N_H)^2
\label{eq:convfact}
\end{equation}

Where the units are ergs cm$^{-2}$. The $1\sigma$ dispersion of the
data-points around this relation is  $0.1$ dex, which can be taken as
an upper limit to the error on the source flux introduced by ignoring
the spectral shape of faint sources. From the observed count-rates we
thus estimated fluxes for 127 X-ray sources with $<$50 counts and with
$N_H$ estimates coming from either the $A_V$ or the JHK photometry. 
The gray histogram in figure \ref{fig:cf} shows the distribution of
$N_H$ for these sources. Results are reported in table \ref{tab:xsrc}.
Note that we do not quote X-ray fluxes for faint sources without an
independent extinction estimate. Upper limit fluxes for 425 X-ray
undetected stars in the ACIS FOV and with an absorption estimate (out
of a total of 1463) were similarly computed from the upper-limits to
the count rates. 

\section{Results - Membership}
\label{sect:memb}

In this section we define a sample of 491 likely NGC~2264 members
within the FOV of our ACIS observation. X-ray sources account for 408
objects while the remaining 83 are selected on the basis of the
strength of $H_\alpha$ line and of optical variability as described in
\S \ref{sect:starchar}. The likely members are indicated in table
\ref{tab:master} by an asterisk following the identification number. We
first consider the possibility of extragalactic contamination and then
discuss separately the nature of X-ray sources with and without
optical/NIR counterparts.

\subsection{Extragalactic X-ray sources}
\label{sect:extrag}

Extragalactic contamination is expected only among X-ray sources that
lack an optical and/or NIR counterpart. We reach this conclusion by
considering the 489 ACIS sources of likely non stellar origin detected
in the {\em Chandra Deep Field North}  \citep[CDFN;
][]{ale03,bar03}\footnote{We excluded 14 sources identified as stars by
\citet{bar03}.}. After defining, for each CDFN source, random positions
in our FOV, we compared their observed count rates with upper limits
computed from our ACIS data at those positions (see \S
\ref{sect:crossid}). We then selected CDFN sources that would have been
detected with our exposure: a total of $139\pm 3$ objects\footnote{Mean
value and 1$\sigma$ uncertainty result from repeating the experiment 10
times, varying the position of simulated extragalactic sources.}. Next,
we compared the positions of these simulated CDFN AGNs in the optical
(I vs. R-I) and NIR (H vs. H-K) color-magnitude diagrams\footnote{The
NIR photometry of the CDFN sources was taken from 2Mass.} with those 
of the X-ray sources in the NGC~2264 exposure that could be placed in
the same diagrams. These plots show that AGNs are on average
considerably fainter than our identified X-ray sources that can be
placed in either of these diagrams and only two or three of the AGNs
that we could have detected occupy positions that overlap with the loci
where our X-ray sources are found. We note moreover that in this
comparison we have totally neglected the effect of extinction. Because
of the dark cloud in our line of sight, the number of AGNs we are
sensible to is surely much reduced with respect to the above
estimate. Moreover the optical and NIR luminosities of AGNs would also
be considerably reduced. We therefore conclude that the contamination
of AGNs to the sample of X-ray selected likely members with optical/NIR
identification is negligible.

\subsection{X-ray sources with optical/NIR counterparts}
\label{sect:membcount}

In figure \ref{fig:cmdO} we showed the optical CMD for the 300 X-ray
sources that can be placed in such a diagram and for the 83 other
undetected likely members discussed in \S \ref{sect:starchar}. Note
that the X-ray sources (filled symbols) are for the most part found in
the locus expected for NGC~2264 members (i.e. the previously defined
cluster locus). However some X-ray sources lie below that locus, their
position being compatible with that expected for MS foreground stars.
Very few, if any, of the X-ray sources in this diagram may be
background MS stars, which are expected to lie below the 1Gyr
isochrone. In order to reduce contamination in the sample of likely
members to a minimum, we exclude X-ray sources that lie outside the
cluster locus, i.e. below $10^{7.1}$ Myr isochrone and to the right of
the 0.8$M_\odot$ evolutionary track. We make an exception for two
sources: \#305 (I=17.28, R-I=0.98), associated with the IRAS source
IRS-1, classified as a Class I object by \citet{marg89} but more
recently suggested to be a deeply embedded B star in a more evolved
evolutionary stage \citep{schr03}\footnote{The ACIS spectrum indicates
$N_H=4.8^{8.9}_{3.2}\ 10^{22}$ (90\% confidence interval),
corresponding to $A_V= 20-56$.}; \#309 (I=16.45, R-I=0.76), a known
member, being indeed the peculiar and well studied accreting
binary system KH~15D \citep[e.g.][]{ham05,herb05,dahm05}. The 8
sources that lie to the right of the 0.1$M_\odot$ track are considered
likely members, either of very low (sub)stellar mass or very absorbed.
Lacking $A_V$ values from optical spectra, we can check these two
possibilities using our X-ray data and the JHK photometry. Out of these
8 sources, 4 (\#125, \#192, \#251, \#273) are highly absorbed as
indicated by their X-ray spectra (or hardness ratios) and/or their
positions in the J-H vs. H-K diagram. Other three sources (\#215,
\#316, \#358) however appear to have negligible absorption and are good
candidates for detected brown dwarfs. 

\begin{figure}[t]
\centering
\includegraphics[width=8.7cm]{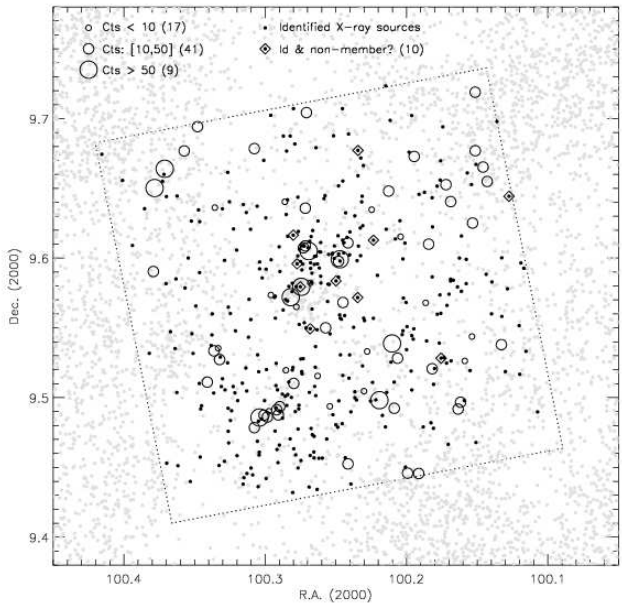}
\caption{Spatial distributions of sources around the ACIS FOV 
(the area within the dotted square). Light gray points:
cataloged optical/NIR objects not detected in X-rays. Black
symbols: X-ray detected objects. Squares with dots: X-ray sources that
might be non-members because of their position in the optical CMD (\S
\ref{sect:memb}). Empty circles with size indicating source statistics
(see legend): unidentified X-ray
sources.}
\label{fig:unid_spatial}
\end{figure}

Only 10 X-ray sources (out of 300) are excluded as members because
incompatible with our cluster locus in the optical CMD and might be
associated with field stars\footnote{\#7, \#51,  \#110,   \#128, 
\#130,   \#171,  \#224,  \#248,   \#258, and \#267.}. Among them only
one (source \#258, an X-ray faint $L_X/L_{bol}=-4.8$ and soft G5 star)
could be placed in the HR diagram of figure \ref{fig:HR}, and is there
placed on the 100Myr isochrone.  Although we exclude these 10 sources
from our sample of likely members, some of them might actually be
members: i) in figure \ref{fig:cmdO} six very likely members, 5
undetected and one detected CTTSs, are found below the cluster locus.
This indicates that some members do have position in the optical CMD
that differ from the expectations; ii) the spatial distribution of
these 10 sources (figure \ref{fig:unid_spatial}, squares+dot symbols)
is similar to that of cluster members, and not uniform as expected from
foreground stars; iii) the ACIS spectra and hardness ratios indicate
that 5 of these sources suffer high extinction in the X-ray band, which
is incompatible with foreground MS star (but not with background
giants).

We now statistically estimate the number of foreground stars that,
present in our FOV and detected by ACIS, contaminate our list of 290
(=300-10) X-ray detected cluster-locus candidate members, as well as
the 10 X-ray sources outside of the cluster locus. We considered the
volume-limited stellar samples of the NEXXUS survey \citep{sch04},
extracting from it a near-complete sample of stars of spectral types F
to M within 6 pc from the Sun. These stars are optically well
characterized and have known distances and X-ray luminosities as
measured with {\em ROSAT}. We estimated the number of foreground stars,
$N_{fg}$, in our FOV assuming that the spatial density of field stars
in the direction of NGC~2264 is uniform and equal to that in the solar
neighborhood\footnote{This is justified by the low equatorial longitude
of NGC~2264: $2.2\deg$.}. We next randomly drew $N_{fg}$ stars from the
NEXXUS sample and assigned them random distances, $d$, in the range
[0-760]pc, according to an uniform stellar density. With these
distances and assuming an isothermal X-ray spectrum with kT=0.3 keV, 
we converted the NEXXUS X-ray luminosities to ACIS count rates using an
$N_H$-dependent conversion factor calculated with the ``Portable,
Interactive Multi-Mission Simulator'' (PIMMS).  The hydrogen column
density toward each star was estimated as $\rho\times d$, where the
volume density $\rho$ was set to 0.3\ $cm^{-3}$ and the distance is
expressed in cm\footnote{With the chosen value of $\rho$, taking
$N_H/A_V=1.6\ 10^{21}$ and d=760pc (the assumed distance of NGC~2264),
we obtain $A_V$=0.44, in agreement with the median of values measured 
for NGC~2264 members.}. We finally compared these simulated
count-rates with upper limits estimated at random positions in the ACIS
FOV estimated using PWDetect (\S \ref{sect:crossid}). This simulation
was repeated 100 times, varying the sample of NEXXUS stars and their
position in space. On average, the sample of foreground stars that we
would have detected counts 20 stars, 15 of which should fall in the
cluster locus and 5 outside of it. About half of the 10 stars outside
the cluster locus are thus expected to be cluster members, thus
reinforcing the arguments given in the previous paragraph. Within the
cluster locus, only 5\% (15/290) of the X-ray sources are estimated to
be non-members\footnote{The fraction of non-members is expected to be
higher among X-ray sources with $M/M_\odot > 1$ and ages $\sim
10^7$yr.}. Given that eighty of the X-ray sources in the cluster locus
had not been previously selected as likely members because of either
their $H_\alpha$ emission or optical variability, we estimate that our
sample contains $\sim65$ new optically visible members.

Fifty-one more X-ray sources were identified with optical or NIR
objects but could not be placed in the optical CMD (14 have I
magnitudes but no R, 37 are only detected in 2MASS). Extending the
previous result and because we can exclude an extragalactic nature for
these sources, we consider these stars as additional candidate members.
Only four of them  had previous indication of membership from
$H_\alpha$ or optical variability.


\subsection{Sources with no optical/NIR counterpart}

Sixty-seven ACIS sources are not identified with any object listed in
the full-field optical/NIR catalogs we have considered\footnote{Four of
these X-ray sources were actually identified with MIR and/or mm sources
in \S \ref{sect:crid_mir}. Because these catalogs are not spatially
complete, for uniformity we here treat them as unidentified.}. None of
them was detected by \citet{fla00} with the ROSAT HRI\footnote{In table
\ref{tab:master}, due to the large uncertainty of the HRI positions,
our ACIS source \#\ 376 is associated with HRI source 138; the HRI
source is however closer to the brighter ACIS source \#375, with which
it is most likely associated.}. With respect to identified sources,
they have significantly fewer counts. Fifteen non identified sources
have detection significance $<5.0\sigma$ and the 10 expected spurious
detections (cf. \S \ref{sect:srcdet}) will be most likely found in this
group. Nine have more than 50 counts and were subject to spectral
analysis.

As discussed in \S \ref{sect:extrag}, depending on the optical depth of
the background cloud, a number of AGNs are expected to be detected in
our FOV and will be found among the non-identified sources. It is
therefore reasonable to ask whether the characteristics of our sources
without counterparts are compatible with an AGN nature. The X-ray
spectra of AGNs should be rather hard and well fit by power law models
with indexes between $\sim$0.9 and $\sim$1.9\footnote{80\% interval of
the power indexes reported by \citet{ale03} for the sources in the CDFN
that we would have detected in our ACIS data and for which the
1$\sigma$ uncertainty in the index is lower than 0.1.}. Their
lightcurves should be constant or slowly varying. They should be
distributed uniformly in space, or anti-correlated with the cloud
optical depth. With respect to this latter point, figure 
\ref{fig:unid_spatial} shows the spatial distribution of several
classes of objects: circles of three different sizes indicate X-ray
sources without counterparts in three source count ranges. Other X-ray
sources are shown with black dots, while all the other objects in our
master catalog are shown in gray. These latter are for the most part
background field stars and their density is a good indicator of the
optical depth of the molecular cloud. Note how the X-ray sources with
counterparts, i.e. likely members, lie preferentially in front of or
close to the cloud and have an highly structured distribution
\citep[c.f., ][]{lamm04}, with at least two concentrations in the south
and toward the field center, corresponding to two well known embedded
sub-clusters roughly centered on IRS~1 \citep[Allen's source,
e.g.][]{schr03}  and IRS~2 respectively (e.g. \citealt{will02}).

\begin{figure}[t]
\centering
\includegraphics[width=8cm, bb= 0 420 610 660]{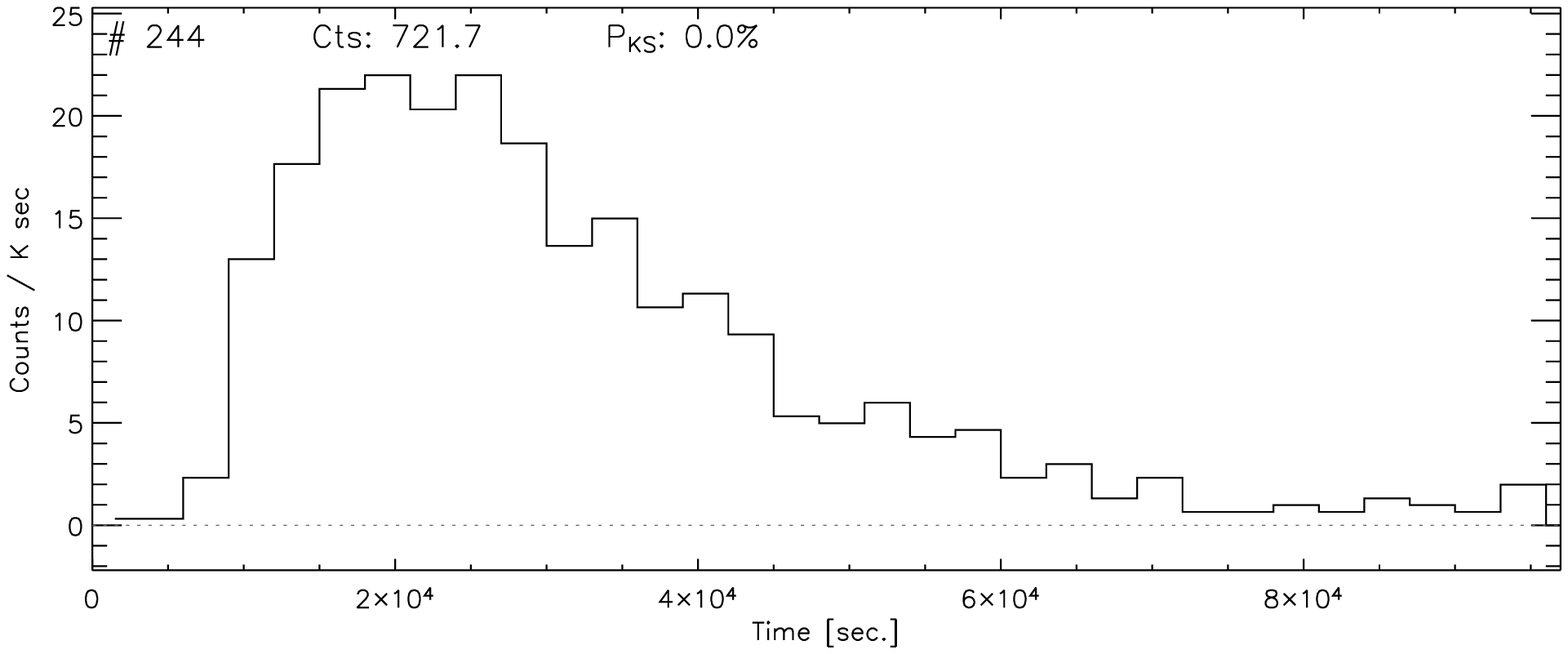}
\includegraphics[width=8cm, bb= 0 420 610 660]{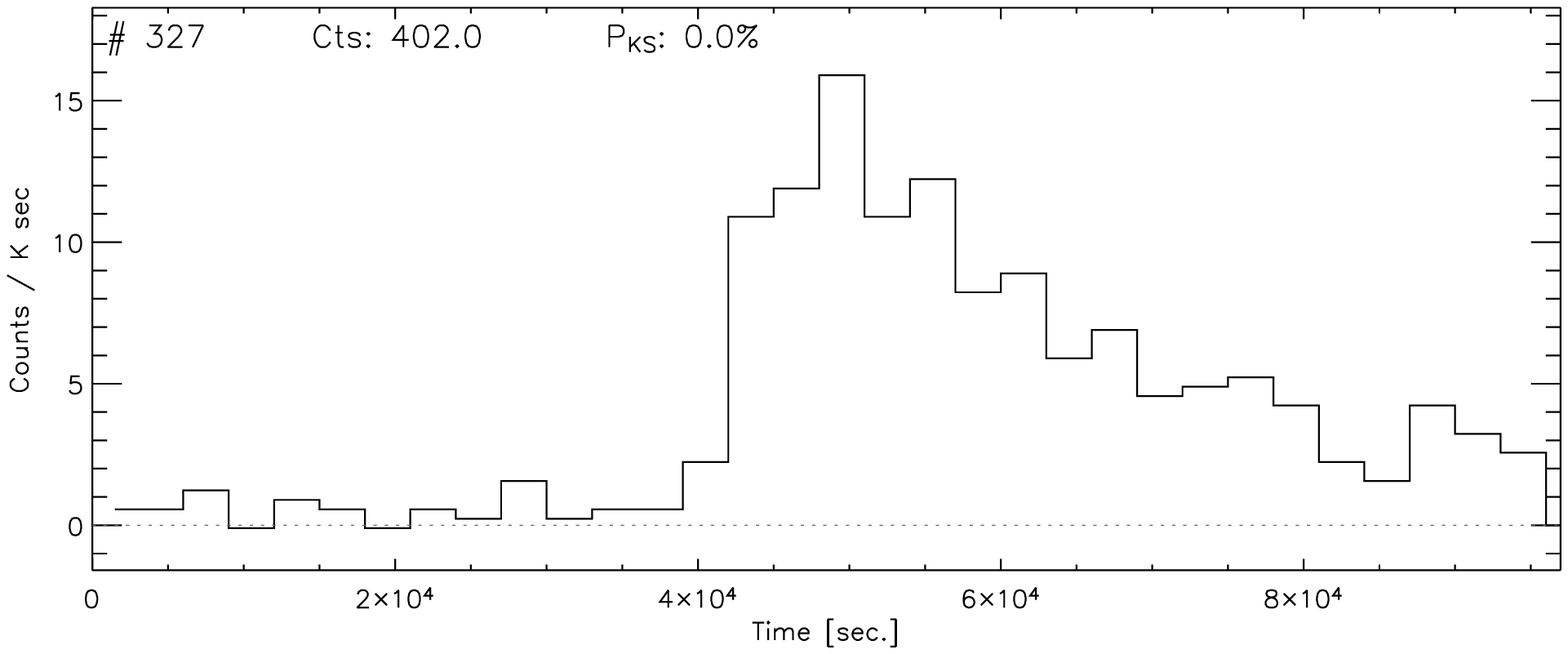}
\includegraphics[width=8cm, bb= 0 420 610 660]{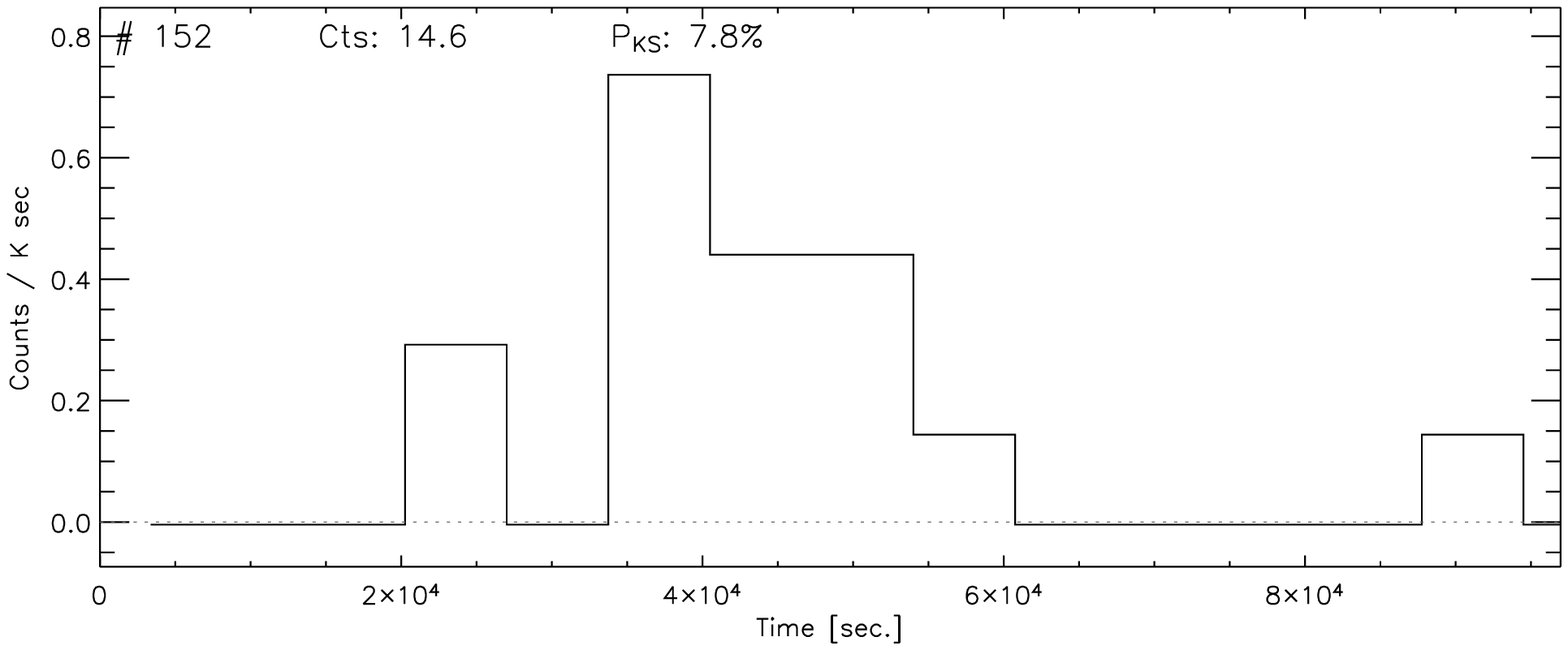}
\includegraphics[width=8cm, bb= 0 420 610 660]{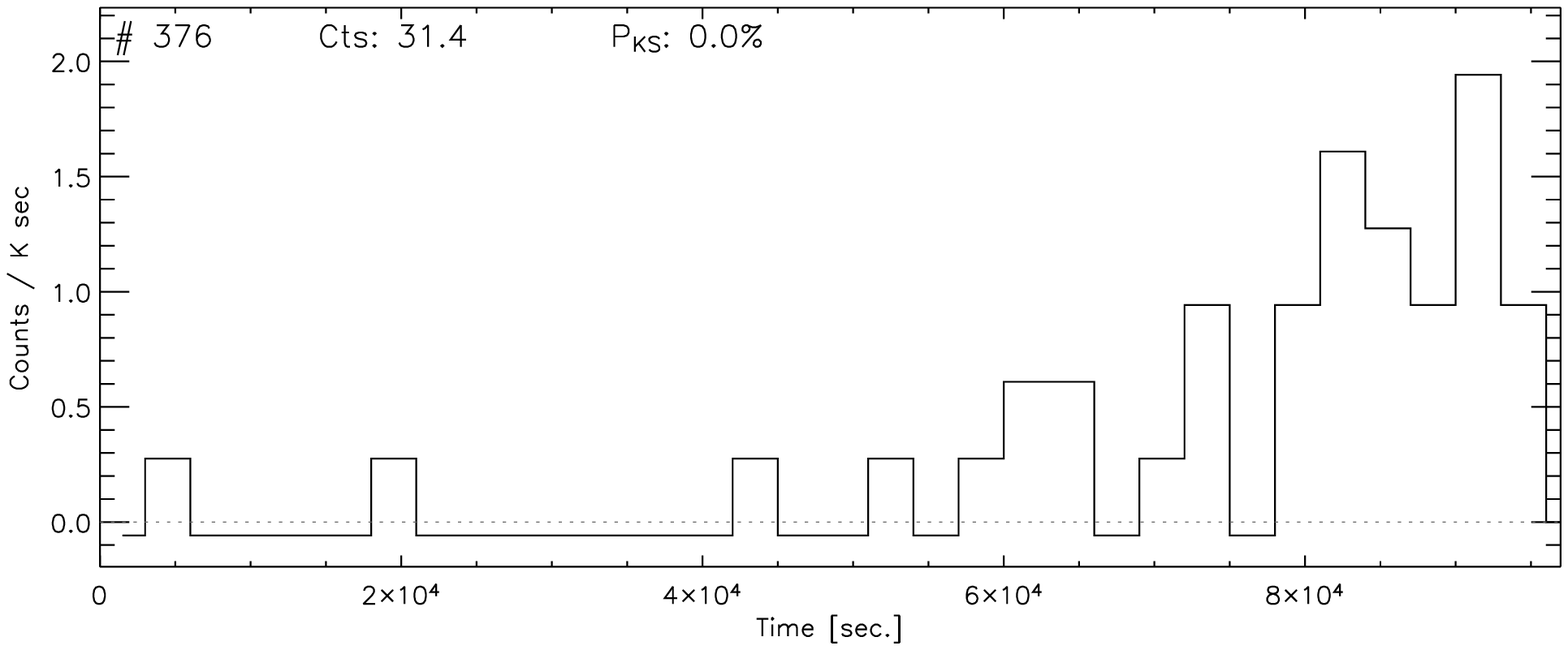}
\caption{Lightcurves for four ``flaring'' sources (\#152, \#244, \#327 and
\#376) with no optical/IR
counterpart. The background subtracted count rate is plotted vs. the
time since the beginning of the observation. Source number, net counts
and the result of the KS variability test are given in the upper part of
each panel.}
\label{fig:flares}
\end{figure}

We first discuss the nine unidentified sources with more than 50
counts: two of them (\#244, \#327), located in the IRS~1 and IRS~2
regions, show distinct long lasting flares (fig. \ref{fig:flares}) and
are therefore most likely PMS stars or YSOs. As discussed in \S
\ref{sect:crid_mir}, source \#244 is actually identified with a {\em
SPITZER} source and, although missing in the 2MASS catalog, it is
rather bright in K.  The spectra of the other seven are compatible with
both isothermal models and power law models. Assuming isothermal
models, they have, with respect to the 190 sources with counterparts
and $>$50 counts, higher extinctions and temperatures (c.f. fig.
\ref{fig:kt1_nh}): the median $N_H$ is 2.7 $10^{22}\ cm^{-2}$, vs. $7\
10^{20}\ cm^{-2}$ and the median $kT$ is 13.7 keV vs. 1.33 keV. Neither
the $N_H$ nor the $kT$ are incompatible with those of other highly
extinct X-ray sources with counterparts (cf. fig. \ref{fig:kt1_nh}),
which we have argued are most likely {\em not} AGNs. Due to the high
extinction, the $L_X$ of these sources (if at the distance of NGC~2264)
is also high: median $\log L_X=30.4$ vs. $\log L_X=29.8$. If instead we
assume the correct models are power laws, the best fit indexes range
between $0.25^{2.0}_{-0.23}$ and $4.1^{6.5}_{2.5}$ (median 1.6). The
two flaring sources have power indexes with 90\% confidence above the
range expected for AGNs and the same is true also for source \#\ 162
($\Gamma = 4.1^{6.5}_{2.5}$), which also lies in the IRS~2 region and
is therefore likely to be a star (the isothermal fit yield a rather
common $kT=1.27$keV).  Two more (\#~228, \#~274) of the remaining six
objects, although with rather hard and absorbed spectra, are clustered
around IRS~2 and are therefore also likely to be YSOs. We thus conclude
that 50\% (5 out of 9) or more of the unidentified sources with more
than 50 counts are likely to be stars. 

Turning to the remaining 58 sources with less than 50 counts, two show
flares (\#152 and \#376, fig. \ref{fig:flares}) and are likely to be
stars. The others, although maybe a little less spatially concentrated
than sources with counterparts, seem to follow a similar distribution
in the sky. We conclude that, rather qualitatively, a sizable fraction
(of the order of $50\%$) of them are associated with NGC~2264. This
conclusion is corroborated by the distribution of unidentified sources
in the $HR_1$ vs. $HR_2$ hardness ratio diagram, figure
\ref{fig:hr1_hr2}. We show for reference a grid of expected loci for
absorbed isothermal spectra and the region where power law sources with
indexes between 0.9 and 1.9 should lie. Both grids are computed using
PIMMS. We note that both $HR_1$ (most sensitive to $N_H$) and $HR_2$
(most sensitive to kT) are on average significantly different for
sources with and without counterparts, indicating that non-identified
sources are on average characterized by hotter and more absorbed
emission. However, the region occupied by about half of the
unidentified sources is also occupied by a number of absorbed sources
with counterparts, i.e. likely members, while the rest appears to be
significantly hotter and possibly compatible with the expected AGN
locus. 

In conclusion, our 67 sources lacking optical/NIR counterparts are good
candidates for new embedded members. Given their absorption these are
rather luminous X-ray sources. They are thus unlikely to be low mass
stars that have escaped optical/NIR detection because intrinsically
fainter than the detection limit. Given the small dependency of
$L_X/L_{bol}$ on mass \citep{fla03a,pre05}, low mass stars are indeed
usually faint in X-rays. Non-identified sources might be embedded
protostars (class I and class 0 objects), medium/high mass very
obscured PMS members of NGC~2264 or extragalactic objects shining
through the background molecular cloud. They certainly deserve to be
followed up with more sensitive IR observations.

\section{Results - X-ray activity}
\label{sect:results}

Our data indicate that the source of X-ray emission in NGC~2264 low
mass members is hot (0.3-10 keV) thermal plasma. The X-ray emission is
highly variable in time, the most prominent phenomena being impulsive
flares due to magnetic reconnection events. These observations fit well
with a solar-like picture of coronal emission and are quite usual for
PMS stars. They are, for example, in qualitative agreement with those
recently reported for the $\sim$1Myr old stars in the ONC by the COUP
collaboration \citep{pre05}. The spectral and temporal characteristics
of PMS stars are, broadly speaking, also similar to those of active MS
stars, e.g. in the young Pleiades cluster. The most striking
differences with respect to MS stars are the X-ray luminosities and the
plasma temperatures, both of which are usually found to be higher. As
for $L_X$ we note that the fractional X-ray luminosities
($L_X/L_{bol}$) are, in PMS stars, comparable or smaller than those of
saturated MS stars. Therefore, the higher $L_X$ can be explained by the
almost saturated emission of PMS stars and by their larger bolometric
luminosities. The higher temperatures might instead indicate a
difference in the heating mechanism and/or, a larger contribution to
the average flux of flares with hard spectra. At least three questions
remain open. First and foremost, the nature of the ultimate mechanism
that sustains PMS coronae. While for partially convective MS stars this
is identified with the $\alpha-\omega$ dynamo thanks to the observed
relation between activity and stellar rotation, no such evidence is
available for PMS stars \citep{fla03a,pre05,reb06}. Second, the extent
and geometry of coronae, and in particular the possibility of
interactions of plasma filled magnetic structures with circumstellar
disks \citep{fav05,jar06}. Third, the role of accretion and outflows:
soft X-ray emission has been inferred to originate both at the
interface between the accretion flow and the photosphere, and within
the stellar jets \citep{kas02,bal03}. In this section we now use our
data to statistically investigate the dependence of activity, in terms
of emission level, variability and spectra, on the stellar and
circumstellar characteristics. First however we discuss more in detail
the results of our X-ray spectral analysis, with respect to plasma
temperatures and absorption values.

\begin{figure}[t]
\centering
\includegraphics[width=8.9cm]{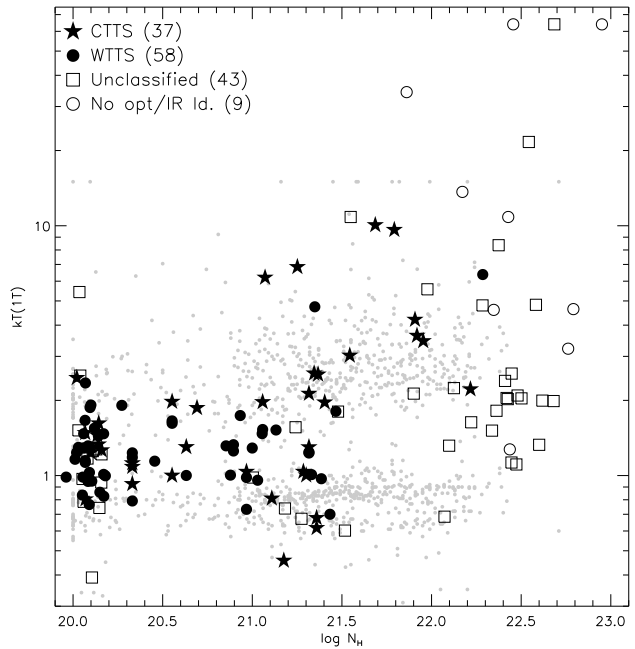}
\caption{Large black symbols: $kT$ vs. $N_H$ for NGC~2264 sources 
for which isothermal models were
adopted. We distinguish (see legend) CTTS and WTTS, 
stars that are identified with
optical and/or NIR sources but for which the PMS class is unknown, and
X-ray sources that are not identified with any known optical/NIR
object. Note that stars with $\log N_H < 20.1$ have been shifted to
that value and their horizontal position slightly randomized so that the
symbols do not overlap completely. 
The small gray circles show for comparison the two temperatures
derived by \citet{get05} for a sample of 566 ONC members observed by the
COUP collaboration.}
\label{fig:kt1_nh}
\end{figure}
\begin{figure}[t] \centering
\includegraphics[width=8.9cm]{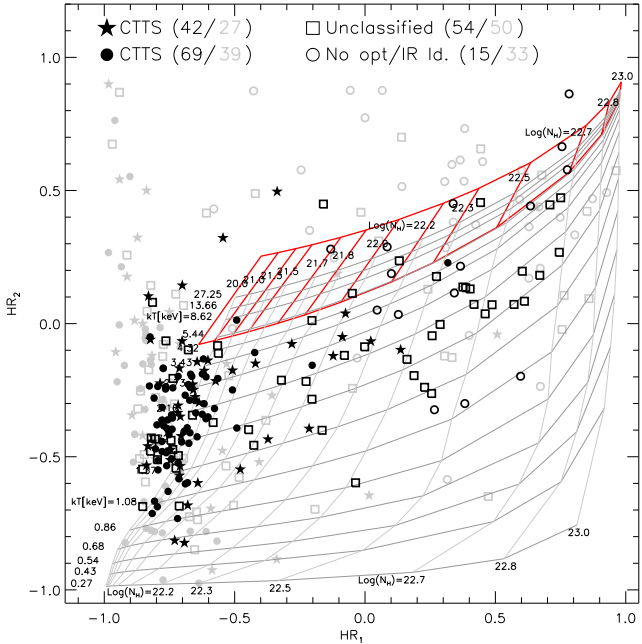}
\caption{$HR_1=(M-S)/(M+S)$ vs. $HR_2=(H-M)/(H+M)$, where H, M and S
are the photon fluxes in the [500:1700]eV, [1700:2800]eV and
[2800:7000]eV bands. Sources for which 1$\sigma$ uncertainties on
$HR_1$ and $HR_2$ are less than 0.3 are plotted in black and those with
larger uncertainties in gray. Symbols are as in Fig. \ref{fig:kt1_nh}.
The number of plotted sources in each class are given in the legend.
The light gray grid indicates the expected loci of isothermal sources
of a given kT  absorbed by interstellar matter of varying $N_H$. 
Values of kTs and $\log N_H$ are given on the left and top/bottom of
the grid. The upper thick-line grid indicates the region expected for
AGNs (power law spectra with indexes between 0.9 and 1.9).}
\label{fig:hr1_hr2} 
\end{figure}
\begin{figure}[t]
\centering
\includegraphics[width=8.9cm]{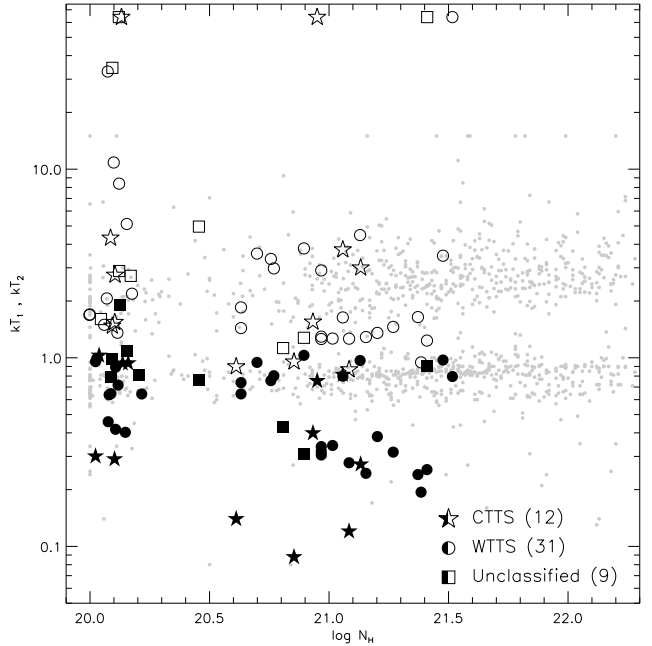}
\caption{$kT_1$ and $kT_2$ as a function of $N_H$ for sources for which 2T
spectral fits were adopted. The results for NGC~2264 presented
in this paper are shown as larger symbols, filled ones for the
cool component temperature and empty ones for the hot component.
Stars, circles and squares indicate CTTS, WTTS and unclassified stars
respectively. The small gray circles show for comparison the two temperatures
derived by \citet{get05} for a sample of 566 ONC members observed by the
COUP collaboration.}
\label{fig:kt12_nh}
\end{figure}

Figure \ref{fig:kt1_nh} shows the plasma temperature as a function of
absorption for all the sources with spectral fits and for which we
adopted isothermal models. Figure \ref{fig:kt12_nh} similarly shows the
two best-fit plasma temperatures of sources for which we adopted 2T
models. First of all we note that 2T models were required only for
$\log N_H < 21.5$, likely because higher column densities absorb the
cool component to the extent that it becomes unobservable. Moreover, in
the $N_H$ range covered by both models, 2T models were statistically
favored in sources with higher statistics while low statistic sources
were in most cases successfully fit with a single plasma component with
temperature roughly intermediate between those of the two components in
2T models. For 1T models we also note a certain correlation between kT
and $N_H$ (fig. \ref{fig:kt1_nh}). While a positive slope of the lower
envelope of the datapoints is easily explained as a selection effect,
this is not the case for the upper envelope. The paucity of sources
with low $N_H$ and high $kT$ indicates that the X-ray emitting plasma
of highly extinct sources is intrinsically hotter than that of
optically revealed PMS stars. These hot, highly extinct sources are
good candidates for embedded Class 0/I protostars, which have already
been suggested to have harder X-ray spectra \citep{ima01}.

\begin{figure}[t]
\centering
\includegraphics[width=8.9cm]{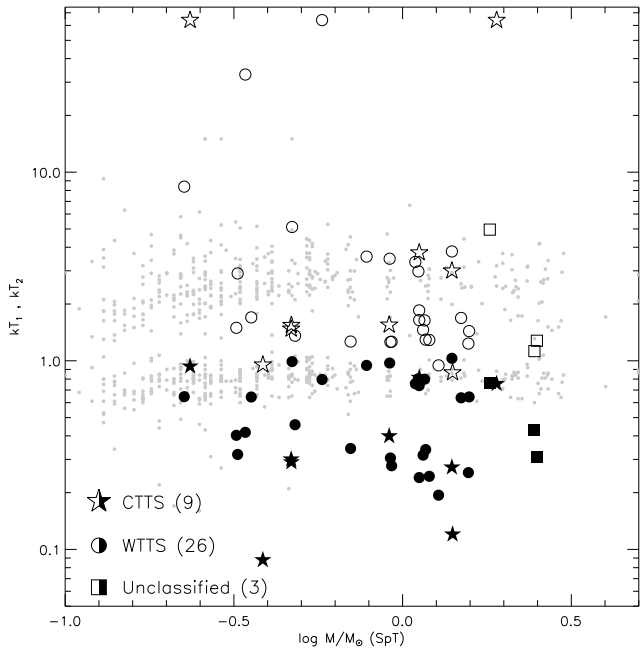}
\caption{$kT_1$ and $kT_2$ as a function
of stellar mass for sources that are both placed in the theoretical HR diagram 
and for
which we adopted a 2T spectral model. Symbols and legend as in 
Figure \ref{fig:m_kt_2t}.
}
\label{fig:m_kt_2t}
\end{figure}

Figure \ref{fig:m_kt_2t} shows, for 2T models, the run of $kT_1$ and
$kT_2$ with stellar mass. Like for the two previously discussed plots,
we also show for reference the temperatures obtained by the COUP
collaboration \citep{pre05} for $\sim$1Myr old stars in the
ONC\footnote{Note that the COUP observations, obtained with {\em
Chandra} ACIS, was 850ksec long. The temperatures are thus derived from
a spectrum that has been integrated in time over a period that is 8.5
times longer than that of our NGC~2264 observation}. With respect to
the ONC, in NGC~2264 temperatures appear to be lower on average. This
could indicate that at $\sim$ 3Myr, the hot flaring component of the
X-ray emission has become less important. It could also result from a
lower fraction of CTTSs given that, as we note below (\S
\ref{sect:actCW}), CTTS tend to have slightly higher plasma
temperatures with respect to WTTS. 

We also note that, with respect to the COUP data, the temperatures that
we find for the two isothermal components show a larger scatter. This
could be due to: i) larger uncertainties in the NGC~2264 results
because of the shorter exposure time and longer distance with respect
to the ONC; ii) the shorter exposure time resulting in a larger
influence of spectral time variability on the time averaged spectra;
iii) a real evolutionary effect (e.g. most ONC stars are CTTS while in
NGC~2264 we observe a more varied mixture of CTTS and WTTS); iv) the
existence of an additional thermal component that is only sometimes
present and/or revealed by the spectral fitting process. In this
respect figures \ref{fig:kt12_nh} and \ref{fig:m_kt_2t} evidence, for
sources fit with 2T models, an interesting feature: an apparent
separation  of the temperatures in two  branches, more evident for the
cool isothermal component. Taking $kT_1=0.5$ keV as the dividing line
between the two branches, we have 23 and 29 sources in the {\em cool}
and {\em hot} branch respectively. The hot branch has median $kT_1$ and
$kT_2$ of 0.8 and 3.6 keV respectively, roughly coincident with the
temperatures of most COUP sources. The cool branch has median $kT_1$
and $kT_2$ of 0.3 and 1.3keV respectively. Noting that this latter
median $kT_2$ is similar to the temperatures found for sources that
were successfully fit by 1T models, it is reasonable to hypothesize
that the EM distributions of these sources have three peaks: one
corresponding to the cool $\sim 0.3$keV component that is not present
and/or visible for sources in the hot branch, and two peaks at hotter
temperatures (i.e. $\sim 0.8$ and $\sim 3.6$keV) that are however well
represented by an isothermal component with intermediate temperatures.
Given the limited statistics of our sources, three component spectral
models, although physically reasonable, are not needed and would
therefore remain unconstrained by the data. 

As for the physical origin of the two branches, we note that with
respect to the hot branch, stars in the cool branch have lower counts
(median counts 270 vs. 630), are significantly less variable according
to the KS test (variability fraction 30\% vs. 66\%, at the 1\%
confidence level) and are slightly more likely to be CTTS (CTTS
fraction: 33\% vs. 23\%). We note that the difference in variability
fractions is unlikely to be explained by statistics alone (c.f. fig
\ref{fig:var_fract_ha} and \ref{fig:var_fract_mass}). If the $kT_2\sim
$3keV component is due to flaring, it is then possible that the
$kT_1\sim 0.3$keV component becomes detectable preferentially when
flaring is absent and therefore the hot component is suppressed. This
soft emission might be physically assimilated to the X-ray emission
from the solar corona. This latter would indeed show temperatures
similar to 0.3keV if analyzed with {\em ASCA} SIS \citep[c.f. Table 1
in ][]{orl01}, an instrument with response similar to ACIS. The
emission measures we derive for the $\sim$0.3keV component of our
NGC~2264 sources, ($3.1\times10^{52}-1.6\times10^{54}$\ cm$^{-3}$)
although much larger than those estimated for the Sun, could still be
explained by much larger filling factors ($\sim1$) and/or larger scale
heights of the densest structures (i.e. active region cores).

\subsection{Activity in CTTS and WTTS}
\label{sect:actCW}

\begin{figure}[t]
\centering
\includegraphics[width=6.4cm]{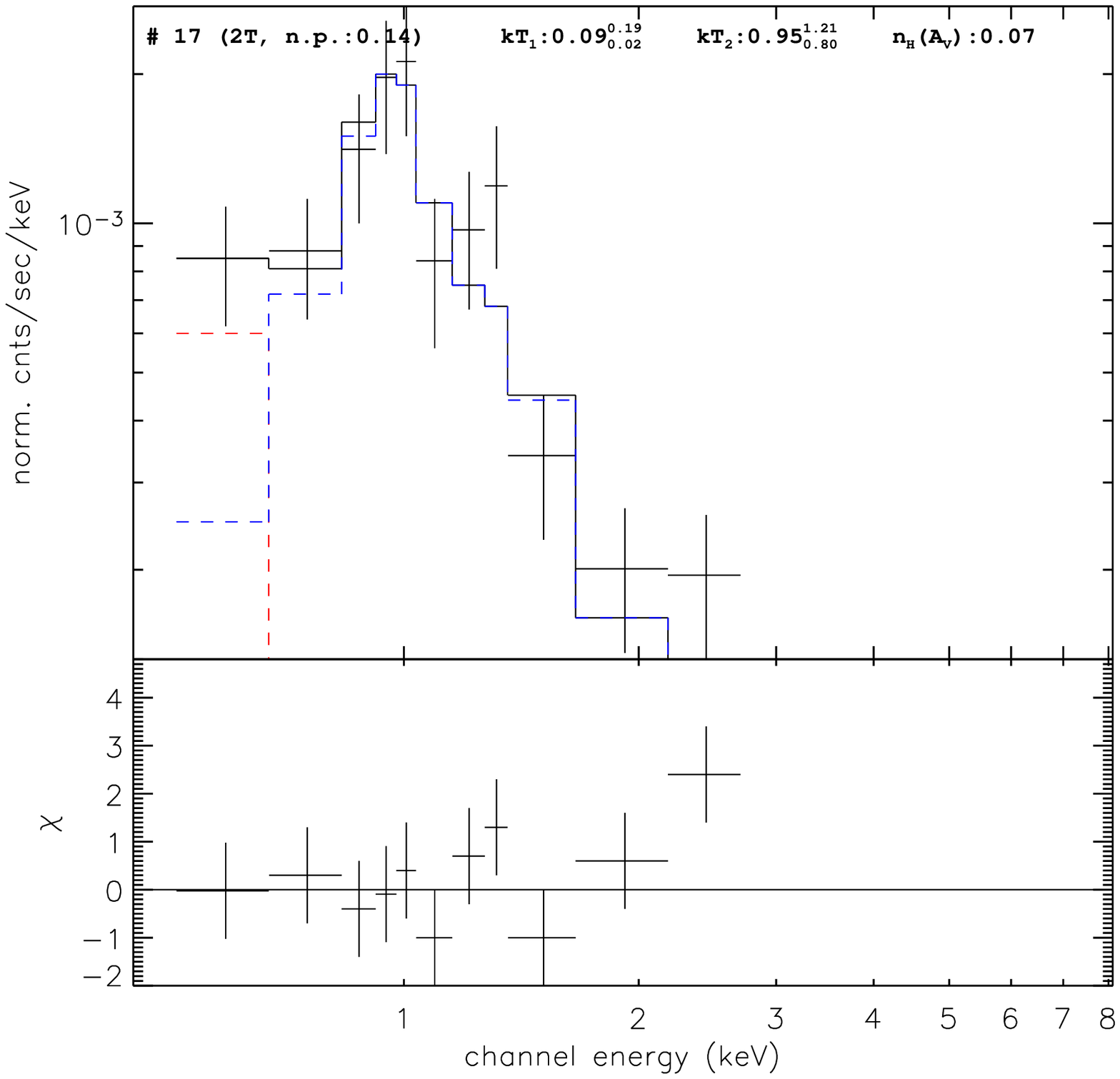}
\includegraphics[width=6.4cm]{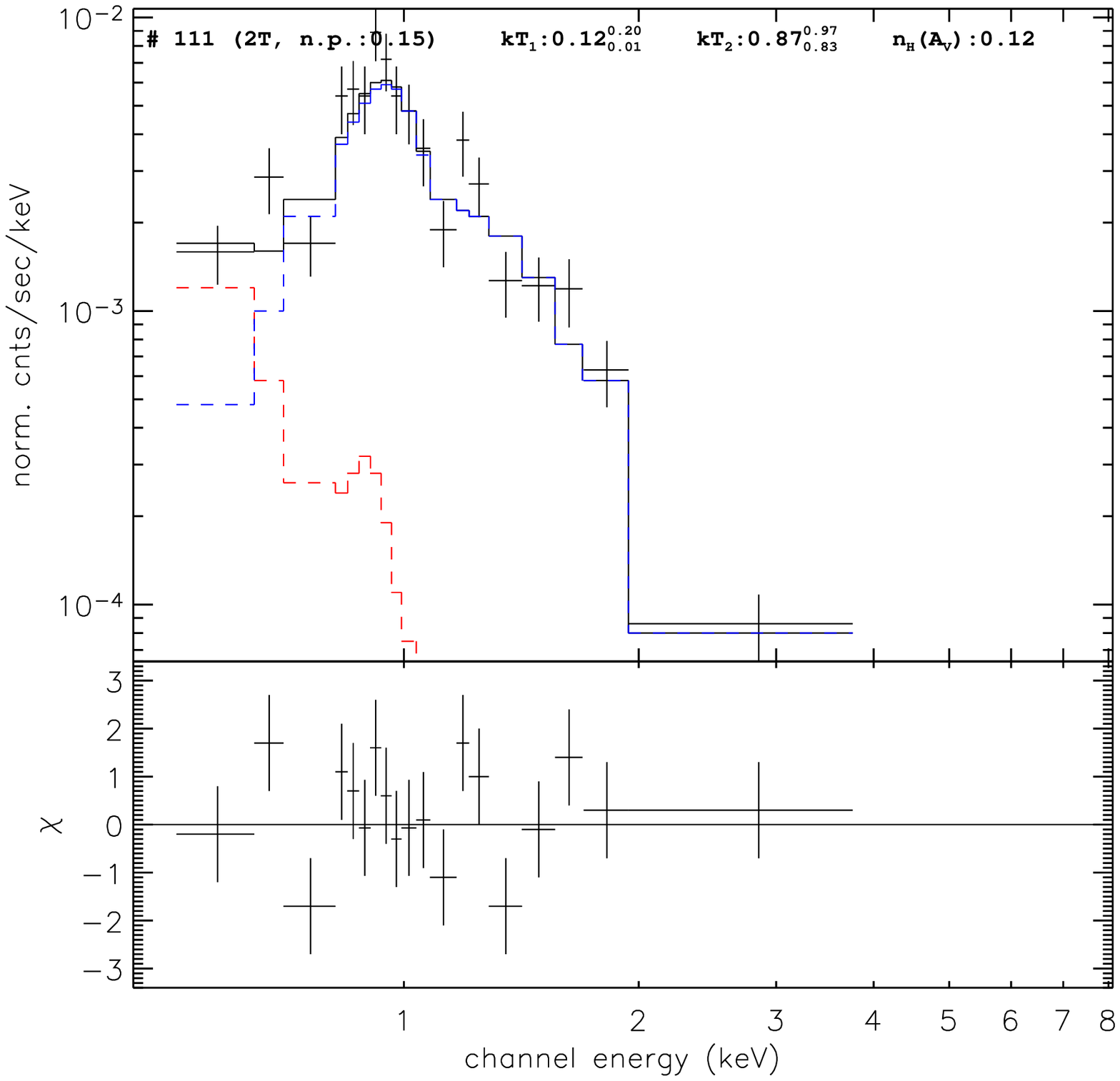}
\includegraphics[width=6.4cm]{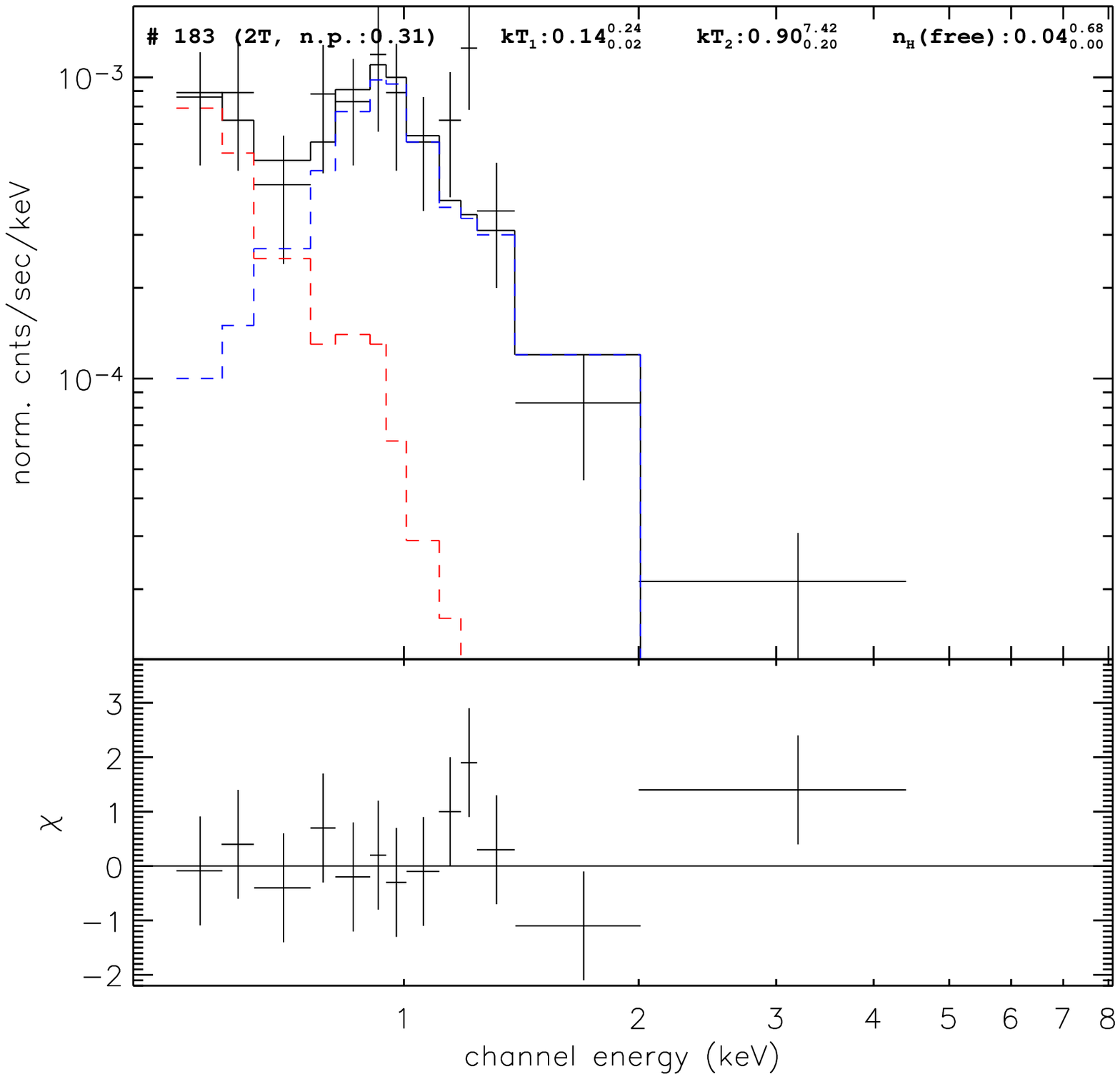}
\caption{ACIS spectra of three CTTS with a very low $kT_1$. At the top
of each panel, we report: source
number, fit model, goodness of fit (null probability), $kT_1$, $kT_2$
and $N_H$ values, in unit of $10^{22}\ cm^{-2}$. For these latter we
give in parenthesis its
origin: ``$A_V$'' or, for source \#183, ``free'' i.e. derived from the 
X-ray spectral fitting.}
\label{fig:low_kt1}
\end{figure}

Comparing the plasma temperatures of CTTS and WTTS that are well fit by
a single temperature model (37 CTTS, 58 WTTS) we learn that the former
are statistically hotter (median: 1.5keV vs. 1.3keV, K-S test
probability that the two distributions are compatible: 0.2\%). A
similar comparison for the two plasma temperatures of stars that
required a 2T spectral model (12 CTTS and 31 WTTS) is inconclusive,
possibly also because of the lower number of stars. However we have
noted above that CTTS appear to be more common among stars with low
values of $kT_1$ (fig. \ref{fig:kt12_nh}). In particular, the three
stars with the lowest $kT_1$ are all CTTS. Figure \ref{fig:low_kt1}
shows the ACIS spectra for these three sources, \#17, \#111 and \#183.
The two thermal components of the best fit model are shown separately
and the temperatures and absorption are reported from table
\ref{tab:xspec}. The most striking case appears to be source \#183,
which has the highest $kT_1$ (0.14keV) and shows a clear double peaked
spectrum. The emission measure ($EM$) of the cool plasma is estimated
to be $5.7\times10^{53}\ cm^{-3}$. The other two sources have lower
$kT$ and higher EMs (\#17: $kT_1=0.09$keV, $EM=2.3\times10^{54}\
cm^{-3}$; \#111: $kT_1=0.12$keV, $EM=1.2\times10^{54}\ cm^{-3}$). We
note also that among the three sources \#183 has the largest $H_\alpha$
line equivalent width: 27.9 vs. 10.9 and 18.2 for \#17 and \#111.

As noted in the previous section, low $kT_1$ values are usually
associated with low $kT_2$. For the three CTTS just discussed for
example the $kT_2$ values, 0.87-0.95keV, are among the lowest observed
and very similar to the cool component for the majority of PMS stars.
This suggests that the ultra-cold component is present and/or
observable with our spectra only when $kT=2-4$kev plasma is absent.

A similar trend can be observed for 1T fits (figure \ref{fig:kt1_nh}):
among the six lowest kTs ($kT < 0.68$keV), the  EW($H_\alpha$) is know
for three stars and in all cases it indicates accretion (i.e. a CTTS).
CTTS thus appear to possess both warmer and cooler plasma than WTTS. If
this results is confirmed with more statistical significance by further
observations, it could imply that the accretion process results in, on
one hand, more frequent/energetic  flares and, on the other hand, a
very cool X-ray plasma produced in the accretion shock (cf. the cases
of TW Hydrae and BP Tau: \citealt{kas02,ste04,sch05}).

We now investigate the X-ray activity levels ($L_X$ and $L_X/L_{bol}$)
as a function of bolometric luminosity, stellar mass and
accretion properties for the subsample of stars for which stellar
masses were derived by placement in the theoretical HR diagram and
interpolation of the SDF tracks (160 X-ray detected stars, excluding
one possible non-member and two stars outside the  tracks). For this
investigation we also include upper limits for 16 X-ray
undetected likely members (empty symbols in figure \ref{fig:HR}). 
Note that a more exhaustive account of the relation between activity
and stellar properties, using the X-ray data presented here in
conjunction with those of \citet{ram04a} for another field in NGC~2264,
and those of  \citet{ram04b} for the ``Orion Flanking Fields'', can be
found in \citet{reb06}. In that paper the  samples for each cluster
included many more stars than we have here, but with on average more
uncertain stellar parameters. Here we take a different approach,
focusing only on the NGC~2264 members in our FOV that are well
characterized optically.

\begin{figure}[t]
\centering
\includegraphics[width=8.9cm]{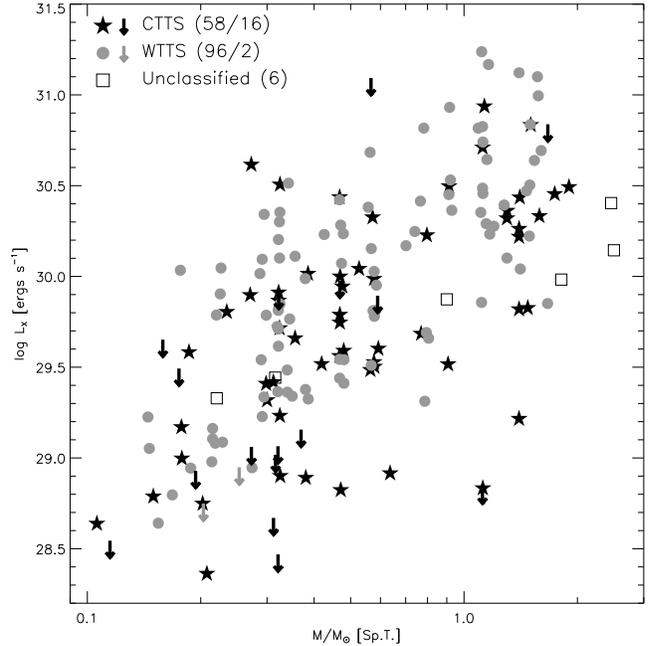}
\caption{log $L_X$ vs. mass for likely members placed in the theoretical
HR diagram. Black stars and arrows indicate CTTSs (detections and upper
limits respectively). Gray circles and arrows indicate WTTSs. Squares
indicate detections of unclassified PMS class. The legend gives the
number of detections
and upper limits for each of the three subsamples.}
\label{fig:lx_mass}
\end{figure}

\begin{figure}[t]
\centering
\includegraphics[width=8.9cm]{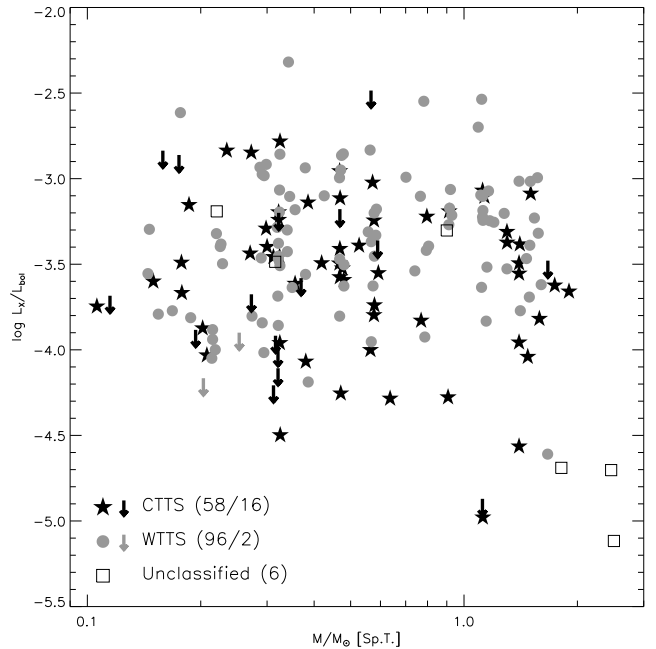}
\caption{log $L_X/L_{bol}$ vs. mass for the same stars of 
figure \ref{fig:lx_mass}.}
\label{fig:lxlb_mass}
\end{figure}

\begin{figure}[t]
\centering
\includegraphics[width=8.9cm]{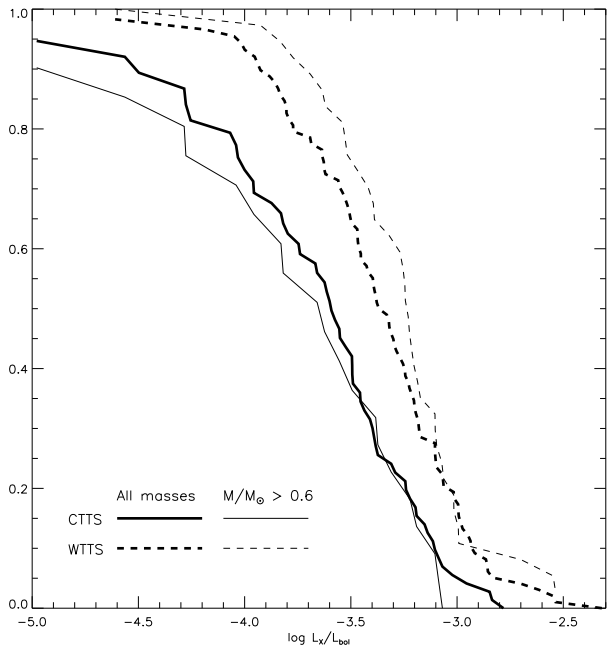}
\caption{Distribution of log $L_X/L_{bol}$ for WTTS and CTTS (solid and
dashed lines, respectively). The thick
lines refer to the whole sample of stars depicted in figure
\ref{fig:lxlb_mass}; the thin lines to stars more massive  than
0.6$M_\odot$. Note how in both cases CTTS are on average less active
than WTTS.}
\label{fig:dist_LxLbol}
\end{figure}

Our $L_X$-mass scatter plot is shown in figure \ref{fig:lx_mass}. We
observe the commonly found mass-$L_X$ correlation \citep[e.g.
][]{reb06}, although with a large spread. The position of upper limits
indicate that our sensitivity limit is $\log L_X \sim 29$ ergs/s, which
appears to correspond to a completeness limit in mass at about
0.3-0.4$M_\odot$. Moreover we note that, at each stellar mass for which
our sample is reasonably complete, CTTS are on average fainter and more
scattered with respect to WTTS, confirming the results obtained for ONC
stars \citep{fla03a,pre05}. A similar plot is shown for $L_X/L_{bol}$
in Figure \ref{fig:lxlb_mass}. $L_X/L_{bol}$ appears to be generally
high, roughly between $10^{-4}$ and the saturation level $\sim
10^{-3.0}$. Twenty-two sources actually have measured $L_X/L_{bol}$
above the saturation limit. However, a large fraction of these, 73\%,
are variable, a significantly higher variability fraction than among
sources below the saturation threshold (16\%). Moreover, most of the
sources with the highest values of $L_X/L_{bol}$ show large flares,
which, if excluded would bring them close or below to the saturation
level. \citet{fla03a} found evidence for ONC stars of a decrease of
$L_X/L_{bol}$ at the very lowest masses. Our sample of stars with mass
estimates is not complete enough at those masses to study this effect
in detail. However, consistently with these results, we do note a
decrease in the upper envelope of the $L_X/L_{bol}$ vs. mass relation
for $\rm M/M_\odot \lesssim 0.3$. Considering CTTS and WTTS separately,
the difference in activity is less striking in this plot with respect
to the $L_X$-mass one. However, figure \ref{fig:dist_LxLbol} shows,
separately for the two PMS classes,  the distributions of
$L_X/L_{bol}$, both for the whole mass range, and for the subsample
with $M>0.6M_\odot$. As noted above, according to the $L_X$-mass
relation, this latter sample should be almost complete.  All the
distributions take into account upper limits via the Kaplan-Meier
estimator. For both samples, and in particular for the mass restricted
one, we observe that CTTS are less active with respect to WTTS.
Median $\log L_X/L_{bol}$ values differ by 0.32 and 0.41 dex for the two
subsamples. The statistical significance of the difference is confirmed
by the five two-sample tests in the ASURV package \citep{fei85}, giving
probabilities that the $\log L_X/L_{bol}$ distributions of CTTS and WTTS
are taken from the same parent distribution $<$0.02\% for the whole
stellar sample and $<$0.3\% for the mass restricted one. We note that
this latter result differs from that of \citet{pre05}, who find, in the
ONC, a statistically significant difference in the activity levels of
CTTS and WTTS only for stars in the 0.2-0.5$M_\odot$ mass range.

Finally we repeat with our data the correlation analysis between
$L_X$ and $L_{bol}$ performed by \citet{pre05} for ONC stars classified
as CTTS and WTTS. Using the estimation maximization (EM) algorithm in
the ASURV package we find very nearly linear correlations between $L_X$
and $L_{bol}$ for the two classes of stars: $\log L_X=
(1.02\pm0.09)\log L_{bol}+ 30.21\pm0.05$ for WTTS ($1\sigma$
dispersion: 0.4 dex) and $\log L_X= (1.02\pm0.12)\log L_{bol}+
29.92\pm0.07$ for CTTS ($1\sigma$ dispersion: 0.5 dex). Similarly to
the ONC case, accreting stars in NGC~2264 thus appear to be on average
fainter than non accreting ones with the same $L_{bol}$ (in this case
by a factor of 2) and to have slightly more scattered $L_X$ values.
However we note that the power-law slope derived by  \citet{pre05} for
accreting stars in the ONC ($0.6\pm0.1$) is significantly shallower
than that derived here and that in the ONC the $1\sigma$ dispersion of
points with respect to the best fit relations appear to be larger: 0.5
vs. 0.4 dex for WTTS and 0.7 vs. 0.5 for CTTS. The differences might be
be interpreted as an evolutionary effect, given that the ONC is younger
than NGC~2264 ($\sim1$ vs. $\sim3$Myr) and that accretion disks are
expected to have evolved significantly in the latter cluster
\citep[][]{fla03c}. We note however that the comparison between the
effects of accretion on X-ray activity in the two regions is made
uncertain by the different accretion indicators used in the two cases,
the $\rm H_\alpha$  and the Ca II equivalent widths for NGC~2264 and
the ONC respectively.

\subsection{KH~15D}
\label{sect:kh15d}

The peculiar binary system KH~15D has been the subject of many
investigations. \citet{herb05} in particular, analyzing the same X-ray
data discussed here conclude that the system is a very weak source of
X-ray emission for its mass and age. They tentatively attribute the low
X-ray emission to the high eccentricity of the binary system and/or to
the close periastron approach of the two stars, that may either disrupt
the stellar magnetosphere and/or adversely affect the stellar dynamo.

The X-ray luminosity we derive for KH~15D (our source \#309), based on
21.6 detected photons, is 1.3 10$^{29}$ lower than the value reported
by \citet{herb05} by 0.07 dex. This difference is small and within the
uncertainties but, given that the two values are derived from the same
X-ray data, we investigated the matter further.  The discrepancy can be
fully explained by the different estimated number of source photons,
21.6 vs. 22.5, the assumed value of interstellar absorption, $9\
10^{20}$ vs. $2\ 10^{20}$ cm$^{-2}$, and the assumed source
spectrum.\footnote{Our $N_H$ is based on $A_V=0.58$, derived here from
published photometry and spectral type. \citet{herb05} derive their
value from $E(B-V)=0.1$. This $E(B-V)$ should  however imply $N_H=5-6\
10^{20}$ cm$^{-2}$ assuming $A_V/E(B-V)=3.1$ and depending on the
assumed $A_V/N_H$ ratio in the range 1.6-2.0 10$^{21}$. As for the
temperature \citet{herb05} assume solar abundances and kT=2.7keV, the
latter based on the uncertain (due to statistics) hardness ratio. The
count to flux conversion factor we used (equation \ref{eq:convfact})
was derived from sources with 50 to 100 counts, which were fit with
APEC models with subsolar abundances and median kT of 1.0 keV. Note
that our harness ratio analysis (c.f. figure \ref{fig:hr1_hr2})
actually indicates a very cool spectrum, kT$\sim 0.3$keV, with large
uncertainties.} The 0.07 dex difference in $L_X$ however is not
particularly relevant for the physical conclusions regarding the low
$L_X$ of the system. KH~15D in not shown in our $L_X$-mass scatter plot
(figure \ref{fig:lx_mass}) because the system is located in the HR
diagram below the grid of the evolutionary tracks (figure \ref{fig:HR})
and we did not derive a mass. Extrapolating the tracks one would
estimate a mass of 0.6-0.7$M_\odot$, consistent with the value used by
\citet{herb05}, 0.6$M_\odot$. Thus placing KH~15D in the $L_X$-mass
diagram we notice that it would fall below the bulk of the other
NGC~2264 members, but in an area that is populated by other CTTSs. The
value of $\log L_X/L_{bol}$ we derive from our data, -3.45, is moreover
perfectly in line with most of the other NGC~2264 members (cf. figure
\ref{fig:lxlb_mass}). We therefore tend to believe that, rather than
being affected by the peculiar binary orbits, the low X-ray emission of
KH~15D is due to the same mechanism that suppresses activity in CTTSs.

\subsection{Embedded XMM-Newton sources studied by \citet{sim05}}
\label{sect:XMM}

\citet[][hereafter SD05]{sim05} studied in detail three embedded
X-ray sources close to IRS~1 (see also \S \ref{sect:crid_mir}) using
{\em XMM-Newton} EPIC data taken in March 2002, i.e. 7 months before
our ACIS exposure. All the three sources are retrieved in our data and
we now compare the results with respect to spectral characteristics,
variability and average X-ray luminosities. 

EPIC source \#26 (our source \#194) was the most stable of the three
sources. The lightcurves were constant in both observations; the
absorption was identical, $N_H=2.6\ 10^{22}$ cm$^{-2}$; and the kT was
also the same within uncertainties\footnote{In this section
uncertainties quoted for quantities derived by SD05 are $1\sigma$,
while for our results we quote 90\% confidence intervals.}:
$2.09\pm0.23$ keV (EPIC) vs. kT=$2.4^{4.8}_{1.5}$ keV (ACIS). The X-ray
luminosity however, when corrected for the different energy band used
by SD05 (1-10keV vs. our 0.5-7keV) and the different assumed distance
(d=800pc, vs. our 760pc) seems to have dropped by a factor $\sim2$
between the two observations.

The EPIC source \#10 (our source \#141) showed a dramatic flare toward
the end of SD05's exposure with the peak count rate reaching $\sim$100
times brighter than the quiescent emission before the flare. Our
lightcurve is instead compatible with constant emission. DS05 analyzed
the spectrum during the flare, while we report results for the average
spectrum, which is however built from only $\sim 65$ photons. There is
no evidence of variation in the absorption: DS05 find $N_H=2.26\pm0.16\
10^{22}$ cm$^{-2}$ vs. our $2.6^{4.3}_{1.5}\ 10^{22}$ cm$^{-2}$. The
temperatures are however very different owing to the bright flare in
the EPIC data: kT=$12.94\pm3.30$ keV vs. kT=$2.0^{6.1}_{1.0}$ keV. The
X-ray luminosities are also very different, with the SD05 value
($L_X=32.37$) about 2 dex larger than the value obtained from our
analysis and corrected for the different bands and distances. It is
therefore likely that we observed the source in a state similar to the
pre-flare state in the SD05 data.

Finally, the lightcurve of EPIC source \#1 (our source \#296) showed,
during the XMM exposure, a remarkable rise in count rate by a factor
$\geq 7$ in about 8 ksec, and then began what appears as a slow decay
for the remaining 30 ksec of the observation. An isothermal spectral
fit gave a very high absorption, $N_H=9.32\pm0.15\ 10^{22}$ cm$^{-2}$,
temperature, kT=$10.38\pm0.54$ keV, and total luminosity $L_X=1.1\
10^{33}$ ergs s$^{-1}$ (in the 1-10keV band). From our ACIS data we
derive a two order of magnitude smaller luminosity $L_X=1.1\ 10^{31}$,
a somewhat colder plasma, kT=$4.8^{9.6}_{3.0}$ keV and a 5-fold lower
absorption $N_H=1.9^{2.4}_{1.6}\ 10^{22}$ cm$^{-2}$. The lightcurve
indicate a roughly linear decay of the count rate during the 100ksec of
our observation from $\sim6$ cts\ ksec$^{-1}$ to $\sim2$ cts\
ksec$^{-1}$. We note that our value of $N_H$ (90\% confidence interval
corresponding to $A_V=10-15$) is roughly in agreement with the
absorption SD05 derived for the source from NIR spectroscopy
($A_V=15-20$) and, contrary to the value observed during the {\em
XMM-Newton} exposure, does not imply a gas-to-dust ratio larger than in
the interstellar medium. We speculate that the high absorption seen by
SD05 is due to a solar-like CME associated with the flare as also
speculated for our source \#71 in \S \ref{sect:spec}.

\section{Summary and conclusions}
\label{sect:sumconc}

We observed NGC~2264 with {\em Chandra}-ACIS for 97\ ksec, detecting a
total of 420 X-ray point sources. We identify 85\% of the X-ray sources
with known optical and NIR objects, while 67 sources remain with no
counterparts in the considered optical/NIR catalogs. More than 90\% of
the 353 X-ray sources with counterparts are expected to be members of
NGC~2264. Using $H_\alpha$ and optical variability data from the
literature we select a further sample of 83 X-ray undetected likely
members in the FOV of our ACIS observation, bringing the census of
likely members with optical/NIR counterparts to 421 stars. Taking
into account the small estimated contamination from field stars, we
have thus increased the known member census of the region by about 100
objects, mostly very low mass stars and including some candidate brown
dwarfs. A further group of 10 X-ray sources, excluded from the member
sample because their position in the optical CMD is discrepant with the
cluster locus, is also likely to include members. Moreover, among the
67 sources with no optical/NIR counterparts, we argue that about half
are previously unrecognized embedded members and good candidates for
X-ray detected Class 0/I sources. The other half is instead likely
associated with extragalactic objects. The coming {\em SPITZER} data,
will be very useful to clarify the nature of each source and will
allow a systematic study of X-ray activity in the protostellar phase.

We determined X-ray unabsorbed fluxes and luminosities for 326 sources
for which absorption could be estimated, either from the X-ray spectra,
from optical spectral types and photometry ($A_V$) or from NIR
photometry. With the aim of shedding light on PMS X-ray activity, we
then performed a detailed study of X-ray lightcurves and spectra, and
studied the relation between the properties of X-ray emitting plasma
and stellar/circumstellar characteristics. We confirm several previous
findings: X-ray luminosity is related to bolometric luminosity and
to stellar mass; $L_X/L_{bol}$ is on average high and rather
independent of mass, other than for a possible drop at $\sim2M_\odot$
and a shallow decrease for $\rm M/M_\odot \lesssim 0.3$. The mass-$L_X$
relation appears to be better defined for WTTS than for CTTS, and CTTS
have on average lower activity levels at any given bolometric
luminosity and mass. We found tentative evidence that CTTS are more
time variable with respect to WTTS, which might be related to the
time-variable nature of the accretion process if this latter plays a
role in the X-ray emission. With respect to spectral characteristics,
the plasma on CTTS is on average slightly hotter than on WTTS, a
finding possibly related to the higher variability of CTTS. However, we
also observe that the sources with the coolest plasma are
preferentially CTTS. Three CTTSs in particular appear to have plasma at
$\sim0.1-0.2$keV, i.e. comparable with the temperatures expected for
plasma heated by accretion shocks, as observed on TW Hydrae
\citep[$kT\sim0.25$keV, ][]{kas02,ste04}. The estimated emission
measures of this cool plasma is between 4 and 17 times larger than on
TW Hydrae ($EM=1.3\times10^{53}$\ cm$^{-3}$), maybe as a result of the
expected larger accretion rates of NGC~2264 stars. These results, taken
as a whole, reinforce the mounting evidence that activity in low mass
PMS stars, while generally similar to that of saturated MS stars, is
significantly affected by mass accretion. This influence has at least
two aspects: accretion is on one hand a positive source of very soft
X-ray emission produced in the accretion shock. On the other hand it 
reduces the average energy output of coronae and makes the emission
more time variable. \citet{pre05} discuss several possible explanations
for the suppression of activity. They favor the idea that accretion
modifies the magnetic field geometry and results in the
``mass-loading'' of field lines, thus hampering the heating of plasma
to X-ray temperatures. It is at the same time conceivable that the
resulting magnetic field structure will be more unstable because of the
the temporal variability of the mass accretion rate as well as the
rotational shear at the inner edge of the circumstellar disk. In order
tackle the so-far elusive activity-accretion relation, a better
characterization of the circumstellar/accretion properties, e.g.
measures of mass accretion rates, is essential. Particularly useful in
these respect would be contemporary observations in the X-ray band and
in accretion/outflow sensitive optical/NIR lines.

\begin{acknowledgements}

The authors wish to tank the anonymous referee that with his
comments helped to improve this work and acknowledge financial support
from the {\em Ministero dell'Istruzione dell'Universit\'a e della
Ricerca}, PRIN-INAF and contract ASI-INAF I/023/05/0.

\end{acknowledgements}

\end{document}